\documentclass[floatfix,aps,prb,showpacs,twocolumn,amsmath,preprintnumbers,nofootinbib]{revtex4-1} 

\usepackage{epsfig,verbatim}
\usepackage{amssymb,amsfonts}
\usepackage{graphicx}
\usepackage[caption=false]{subfig}

\def\art{paper }
\def\jrn#1#2#3#4#5#6{#3 \textbf{#4}, #5 (#6).} \def\andd{and } %aip

\def\scn#1#2{\section{#1}\lb{#2}}
\def\sscn#1#2{\subsection{#1}\lb{#2}}
\def\bfl{\begin{flushleft}}
\def\efl{\end{flushleft}}
\def\bfr{\begin{flushright}}
\def\efr{\end{flushright}}
\def\bc{\begin{center}}
\def\ec{\end{center}}
\def\be{\begin{equation}}
\def\ee{\end{equation}}
\def\bse{\begin{subequations}}
\def\ese{\end{subequations}}
\def\ba{\begin{eqnarray}}
\def\ea{\end{eqnarray}}
\def\baa#1{\begin{array}{#1}}
\def\eaa{\end{array}}
\def\bw{\begin{widetext}}
\def\ew{\end{widetext}}
\def\nn{\nonumber }
\def\lb#1{\label{#1}}
\def\bit{\begin{itemize}}
\def\eit{\end{itemize}}
\def\bco{\begin{comment}} \def\eco{\end{comment}}
\def\bcs{\begin{cases}}
\def\ecs{\end{cases}}

\def\twomat#1#2#3#4{\begin{pmatrix} #1 & #2 \\ #3 & #4 \end{pmatrix}}
\def\twomatsm#1#2#3#4{\left(\begin{smallmatrix} #1 & #2 \\ #3 & #4 \end{smallmatrix}\right)}
\def\twocol#1#2{\begin{pmatrix} #1 \\ #2\end{pmatrix}}
\def\threecol#1#2#3{\begin{pmatrix} #1\\#2\\#3 \end{pmatrix}}

\def\fourcol#1#2#3#4{\begin{pmatrix}#1\\#2\\#3\\#4\end{pmatrix}}
\def\schrod{Schr\"odinger}

\def\pDer#1{\frac{\partial}{\partial #1}}

\def\Cos#1#2{\, \text{cos}^{#1}\!\left(#2 \right) }

\def\Sin#1#2{\, \text{sin}^{#1}\!\left(#2 \right) }

\def\om{\bar{\omega}}
\def\pcz{\beta_z}

\def\veef{\mathbf{E}}
\def\veeft{\mathbf{E}_{\bot}}
\def\vemf{\mathbf{H}}
\def\vemft{\mathbf{H}_{\bot}}
\def\vena{\boldsymbol{\nabla}}
\def\venat{\boldsymbol{\nabla}_{\bot}}
\def\verad{\mathbf{r}}
\def\vee#1{\mathbf{e}_{#1}}
\def\wavnt{\mathbf{k}_{\bot}}

\def\eigf#1{\mathbf{F}^{(#1)}}
\def\eigv#1{\lambda_{#1}}
\def\evre{R}
\def\evim{I}
\def\tpower{{\cal P}_z}
\def\tpowerbar{\overline{\cal P}_z}
\def\wenbar{\overline{\cal E}}
\def\wen{{\cal E}}
\def\vwenbar{\overline{\cal U}}
\def\vwen{{\cal U}}

\def\t{z}
\def\hbarr{\hbar_\text{w}}
\def\gamman{\Gamma_{\cal N}}
\def\decoO{\hat{\cal G}}

\def\hamo{\hat{h}}
\def\hamto{\hat{\cal H}}
\def\hamod{\hat{\cal D}}

\def\donn{\hat{\rho}}
\def\densnnorm{{\donn^\prime}}
\def\densnnormsq{{\donn^{\prime 2}}}
\def\densnnormOsq{{\donn_0^{\prime 2}}}
\def\densnnormO{{\donn^{\prime (2)}_0}}
\def\densnnormOo{{\donn^{\prime (1)}_0}}

\def\av#1{\langle #1 \rangle}
\def\avo#1{\av{#1}'}
\def\trans#1{\underline{#1}}

\begin{document}

\preprint{\small Phys. Rev. B 94 (2016) 115136 [arXiv:1608.08393]}

\title{
%Density matrix 
Quantum-statistical
approach to 
electromagnetic wave propagation
and dissipation inside
dielectric media
and 
nanophotonic and plasmonic waveguides 
}

\author{Konstantin G. Zloshchastiev}
%\email{k.g.zloschastiev@gmail.com}
\affiliation{Institute of Systems Science, Durban University of Technology, P.O. Box 1334, Durban 4000, South Africa}

%\email{k.g.zloschastiev@gmail.com, kostiantynz@dut.ac.za}

\begin{abstract} 
Quantum-statistical
effects occur during the propagation of electromagnetic (EM)
waves inside the dielectric media or metamaterials, which 
include a large class of nanophotonic and plasmonic
waveguides with dissipation and noise.
Exploiting the formal analogy between the Schr\"odinger equation
and the Maxwell equations for dielectric linear media,
we rigorously 
derive the effective Hamiltonian operator
which describes such propagation.
This operator turns out to be essentially non-Hermitian in general,
and pseudo-Hermitian in some special cases.
Using the density operator approach for general non-Hermitian Hamiltonians, 
we derive a master equation that describes the statistical ensembles of
EM wave modes.
The method also describes the quantum dissipative and decoherence
processes which happen during the wave's propagation,
and, among other things, it reveals 
the conditions that are necessary
to control the energy and information loss inside the above-mentioned materials.
%, which is important for designing the next generation of quantum devices for sensitive and non-invasive measurements.
\end{abstract}

\date{received: 14 March 2016 [PRB], 29 Aug 2016 [arXiv]}
%\date\today


\maketitle

\section{Introduction}

Notwithstanding the long history of studies, 
the propagation of electromagnetic (EM) wave inside
dielectric media remains an important and rapidly
developing topic.
Apart from an obvious theoretical value, it finds numerous applications
in the designs of the nanoscale photonic and plasmonic devices, structures
and metamaterials,
such as lasers, spasers, modulators, waveguides, optical switches, laser-absorbers,
coupled resonators and quantum wells.

During the past decade there has been growing interest
in studying those systems by means of the formal analogy between Maxwell equations in dielectric media and \schrod-type equations,
dubbed here as the Maxwell-Schr\"odinger (MS) map \cite{jbook65,kbbook72,wbook91}.
In this analogy, Maxwell equations are rewritten in the form of the matrix Schr\"odinger equation, 
except that the role of time is played by the coordinate along the direction of wave propagation  
(usually, $z$-coordinate), the Hamiltonian operator is non-Hermitian (NH), 
and the Planck constant is replaced by an effective one \cite{sybook09}.
Therefore, a class of the physical systems that allow such mapping is broadly 
referred as non-Hermitian materials and waveguides.
Moreover, inside this class one can select the subclass of physical systems and phenomena for
which the above-mentioned Hamiltonian operator has real spectrum, in which case it is called
pseudo-Hermitian \cite{ml08,zhu11}.
This pseudo-Hermiticity manifests itself in various phenomena, such as
non-reciprocal light propagation and Bloch oscillations \cite{yz13,fah11,lo09}, 
invisibility and loss-induced transparency \cite{lo11,lre11,fxf13,kcl14,gsd09},
power oscillations \cite{mec08,rme10,rbm12}, 
optical switching \cite{csp92,sxk10,lbd13}, 
coherent perfect absorptions \cite{cgc10,fh12,stl14},
laser-absorbers \cite{lo10,cgs11},
plasmonic waveguides \cite{ad14},
unidirectional tunneling \cite{scg14}, 
loss-free negative refraction \cite{fsa14},
and so on.
These processes can be studied using 
a general theory of pseudo-Hermitian 
(often referred also as ${\cal PT}$-symmetric)
Hamiltonians, which
has originated from works \cite{jd61,sgh92,ben98},
one could mention also the classical results by Dyson \cite{fd56,jof71}.

However, the class of non-Hermitian materials and waveguides is obviously
much larger than its pseudo-Hermitian subclass.
Indeed, as a result of the interaction of EM waves with their environment (which
can be very diverse and uncontrollable),
the description of their propagation
requires the usage of the NH Hamiltonians of 
different kinds, not necessarily possessing real eigenvalues.
In other words, this propagation must be described within the framework of a general theory of open quantum systems \cite{bpbook}.
According to that theory, 
for such situations one needs to engage the full description
of the (quantum) statistical ensemble of EM wave modes.
In turn, it requires the usage of the density matrix, instead of a state vector, as a main object of theory.
Therefore, MS map must be used to develop the corresponding generalization, 
which is going to be the main goal of this \art.
Although the density-operator approach for quantum systems driven 
by NH Hamiltonians has been long since known  (see, for instance, the monograph \cite{fbook}),
it has been further developed in the works \cite{sz13,sz14,sz14cor,ser15,sz15,zlo15}.
In the current \art we adapt this formalism 
%\cite{sz13,sz14,sz14cor,ser15,sz15,zlo15}
for the 
purposes of describing the EM wave propagation inside dielectric materials and
waveguides in presence of dissipative effects induced by environment,
as well as for extracting physical information and predicting new phenomena.

The contents of this paper are as follows. 
In section \ref{s-ana}, we provide essential information
about the Maxwell-Schr\"odinger map for EM wave propagation inside dielectric linear media
of a general type. 
We define the appropriate Hilbert space, as well as we introduce
notions and representations, which will be necessary for what follows.
In section \ref{s-nhgen}, we formulate the density matrix approach for NH dynamics adapted for the MS-mapped models and NH waveguides with dissipation
of general type.
We derive a master equation, which governs the statistical behavior
of EM wave modes inside the medium, and describe its properties.   
In section \ref{s-tls}, we 
use the properties of the Hilbert space in our case
in order to introduce the two-level approach to deriving quantum-statistical
observables for any given medium.
This approach reveals more details about physical features of the systems which
are being studied, as well as it explicitly illustrates some statistical 
effects that occur.
Discussions and conclusions are given in section \ref{s-con}.

\scn{Maxwell-Schr\"odinger analogy in dielectric media}{s-ana}

In this section, we formulate the 
formal mapping between  certain classes of Maxwell 
and Schr\"odinger equations.
We rigorously 
derive the effective Hamiltonian operator,
which describes the propagation of EM
waves in dielectric isotropic media along a certain direction.
Due to its generality, the approach is applicable for
studying EM wave propagation inside
a very large class of materials where at least one
preferred direction of propagation can be established.

\sscn{Effective Hamiltonian}{s-ana-ham}

Let us consider EM wave propagating inside a dielectric isotropic linear medium.
For this situation, there are no
free charges and currents,
therefore,
Maxwell equations acquire a simple form:
%(we use Gaussian units):
\bse
\ba
&&
\vena \times \veef
+ 
\frac{1}{c} \pDer t (\mu \vemf)
= 0
,\\&&
\vena \times \vemf
- \frac{1}{c} \pDer t (\varepsilon \veef)
= 0
,\\&&
\vena \cdot (\varepsilon \veef)
= 
\vena \cdot (\mu \vemf)
=
0
,
\ea
\ese
where $\veef = \veef (\verad, t)$ and $\vemf = \vemf (\verad, t)$
are electric and magnetic fields, respectively,
while
the cross and dot denote the vector and scalar
products, respectively.
Here 
$c = 1/\sqrt{\varepsilon_0 \mu_0}$, $\varepsilon_0$ and $\mu_0$
being, respectively, the vacuum permittivity and permeability,
whereas
$\varepsilon$ and $\mu$
are the relative permittivity and permeability
(complex-valued functions of coordinates, in general);
as per usual, one can also express them via
the medium's electric and magnetic susceptibilities: $\varepsilon = 1 + \chi_e$ and $\mu =  1 + \chi_m$.
The electromagnetic Gaussian unit system's conventions can be used here,
as long as 
the physical vacuum is assumed to be fixed in its current state
characterized by the adopted SI values of $\varepsilon_0$ and $\mu_0$.

%and the wave does not cause perceptible perturbations of its dynamics or structure.

Further,
if we align $z$-axis with the direction of wave's propagation 
then, assuming the harmonic time dependence 
of the electric and magnetic fields,
\bse
\ba
&&
\veef (\verad, t) = \veef (x, y, z) \exp{(-i \omega t)}, 
\\&&
\vemf (\verad, t) = \vemf (x, y, z) \exp{(-i \omega t)}
,
\ea
\ese
one can decompose them into the transverse and 
longitudinal (along $z$-axis) components:
$\veef = \veeft + \vee z E_z$,
$\vemf = \vemft + \vee z H_z$,
$\vena = \venat + \vee z \pDer z$,
where 
%$\omega$ is the wavevector in vacuum,
%$\veef$ and $\vemf$ denote the electric and magnetic fields normalized by the vacuum impedance, 
$\vee n$ is the basis vector along the $n$th axis.
One can show that the vectors $\veeft$ and $\vemft$ are essentially two-dimensional:
$\veeft \cdot \vee z = \vemft \cdot \vee z = 0$.
%, where the dot denotes the scalar product.
Correspondingly, Maxwell equations 
%in such a medium
take the form (from now on we adopt the units where $c= 1$)
\bse\lb{e:maxwd}
\ba
&&
\pDer z \veeft 
-
\venat
E_z
+
i \mu \omega (\vee z \times \vemft)
=
0
, \lb{e:maxwd1}
\\&&
\pDer z \vemft 
-
\venat
H_z
-
i \varepsilon \omega (\vee z \times \veeft)
=
0
, \lb{e:maxwd2}
\\&&
\vee z \cdot 
\left( \venat \times \vemft \right)
+
i  \varepsilon \omega E_z
= 0
,
\\&&
\vee z \cdot 
\left( \venat \times \veeft \right)
-
i  \mu \omega H_z
= 0
, \lb{e:maxwd4}
\\&&
\pDer z \left( \varepsilon E_z \right)
+
\venat \cdot \left(\varepsilon \veeft \right)
=
0
,\\&&
\pDer z \left( \mu H_z \right)
+
\venat \cdot \left(\mu \vemft \right)
=
0
,
\ea
\ese
and also for definiteness we assume throughout the paper that the medium is located at 
$\t \geqslant 0$.

\begin{figure}[t]
\begin{center}
\epsfig{figure=
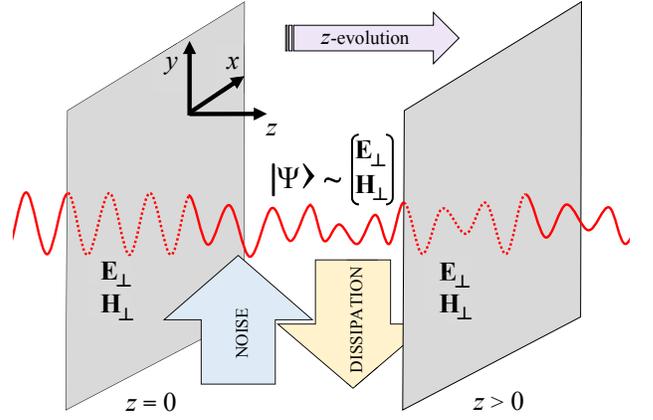,width=0.95\columnwidth,angle=0}
\end{center}
\hspace{0mm}
\caption{Propagation of EM wave through a dielectric medium located at $z \geqslant 0$.
}
\label{f:picgen}
\end{figure}

The system (\ref{e:maxwd})
%-(\ref{e:maxwd4}) 
can be recast in the form,
in which the equations for
longitudinal and transverse vectors
are explicitly separated:
\bse
\ba
&&
i \vee z \times \pDer z \veeft 
=
\hat L_m
\vemft
, \lb{e:maxwo1}
\\&&
i \vee z \times \pDer z \vemft 
=
-
\hat L_e \veeft
, \lb{e:maxwo2}
\\&&
E_z = 
(i \varepsilon \omega)^{-1} \vee z \cdot 
\left( \venat \times \vemft \right)
,
\\&&
H_z = 
- (i \mu \omega)^{-1} \vee z \cdot 
\left( \venat \times \veeft \right)
,
\ea
\ese
where
we denote the following differential operators:
\bse\lb{e:oplem}
\ba
&&
\hat L_e =
\varepsilon \omega - \omega^{-1} \venat \times \mu^{-1}\venat \times
,
\\&&
\hat L_m =
\mu \omega - \omega^{-1} \venat \times \varepsilon^{-1}\venat \times
.
\ea
\ese

Using the 2D property
$\vee z \times \vee z \times = -1$,
the equations (\ref{e:maxwo1}) and (\ref{e:maxwo2}) can be written
in the matrix 
%\schrod-like 
form
\be\lb{e:seprime}
i \pDer z 
%\twocolum{\veeft}{\vemft}
{\veeft\choose{\vemft}}
=
\hamo
{\veeft\choose{\vemft}}
%\twocolum{\veeft}{\vemft}
,
\ee
where we denote the operator
\be\lb{e:hamodef}
\hamo
=
\hat\sigma_2 \hamod
=
\twomat{0}{-\vee z \times \hat L_m}
{\vee z \times \hat L_e}{0}
,
\ee
where
$\hat\sigma_2$ is defined in Appendix \ref{a:op}, and
\be\lb{e:hamoddef}
\hamod
\equiv
\hat\sigma_2 \hamo 
=
\twomat{\hat L_e}{0}
{0}{\hat L_m}
\ee
%where $\hamod$ 
is the auxiliary operator.
% (which is sometimes naively regarded asthe Hamiltonian of the system, due to some similarities between its eigenvalue spectrum and that of $\hamo$).
A schematic drawing of the EM wave propagation as the wave function's evolution along $z$-direction is shown in Fig. \ref{f:picgen}.

\begin{figure}[t]
\centering
\subfloat[Uniform waveguide
% ($\varepsilon$'s and $\mu$'s are constants)
]{
  \includegraphics[width=
	%0.5
	%.88
	0.95
	\columnwidth,angle=0]{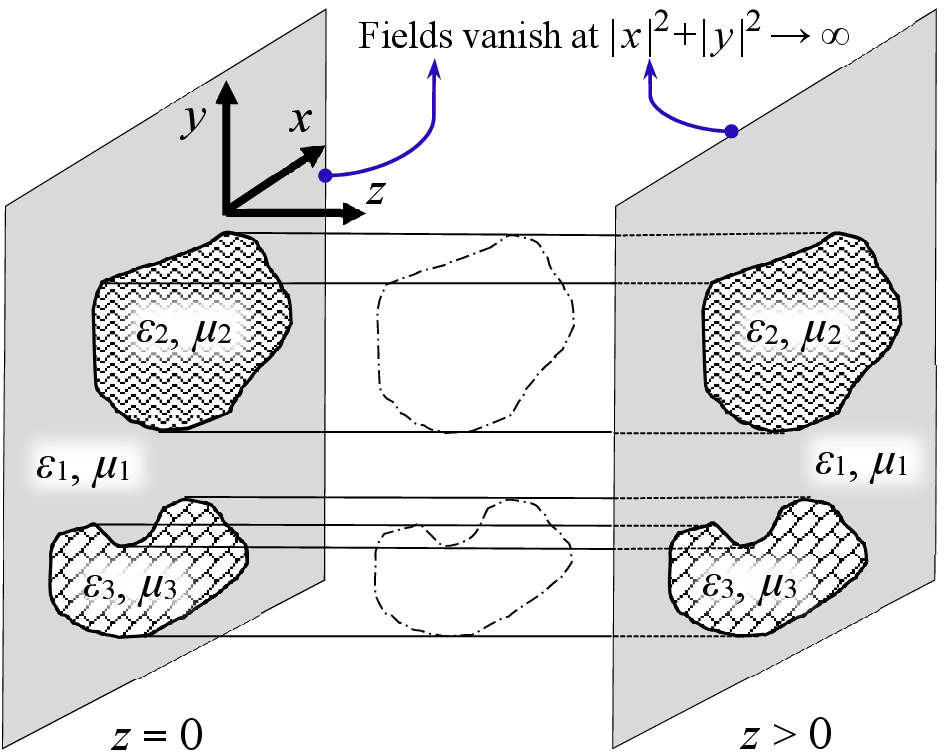}
}
\hspace{0mm}
\subfloat[General waveguide]{
  \includegraphics[width=
	%0.5
	%.88
	.95
	\columnwidth,angle=0]{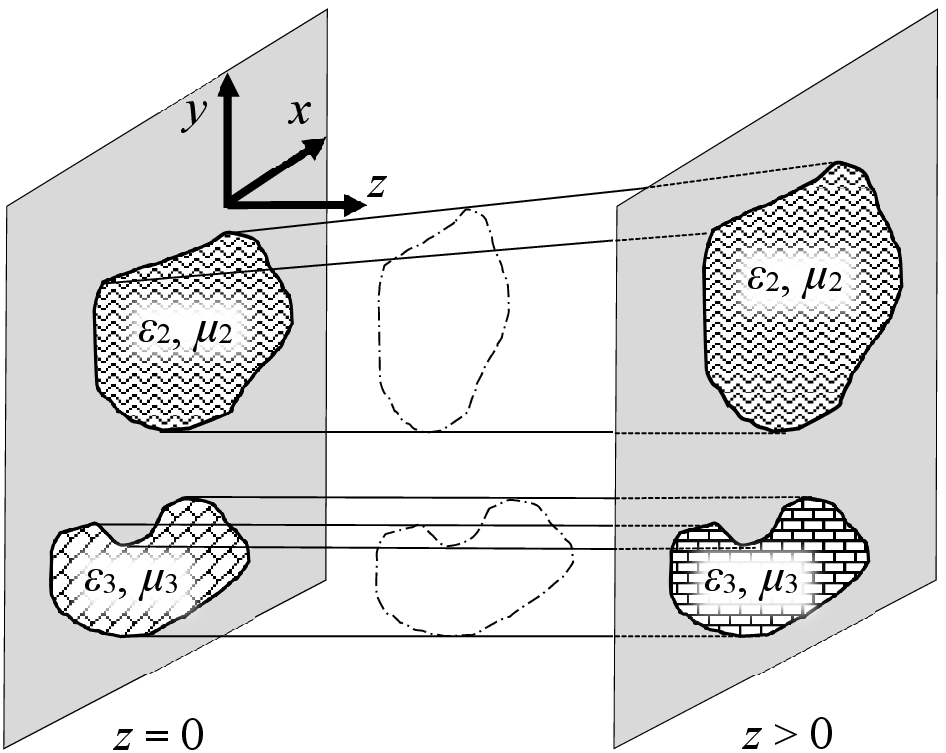}
}
\hspace{0mm}
\caption{EM wave propagation along $z$-direction for different kinds of media.
Vertical planes represent the generalized waveguide's effective cross-section, at different values of $\t$.
For a general medium, deviations from uniformity arise due to varying cross-section (wave-shaded areas on the lower panel) or varying permittivity and/or permeability
% in the internal structure 
(brick-shaded areas on the lower panel).
}
\label{f:picwg}
\end{figure}

Furthermore, using the formulas from Appendix \ref{a:op}, 
one can check that the operator (\ref{e:hamodef}) is non-Hermitian, even 
%in absence of gain or loss (
when both $\varepsilon$ and $\mu$ are real-valued.
This holds for either a uniform waveguide type
geometry [Fig. \ref{f:picwg}a] or for a more general non uniform geometry [Fig.
\ref{f:picwg}b], as long as the issue of transverse fields vanishing at spatial infinity
is properly satisfied, as we will discuss later.
The degree of non-Hermiticity of the theory's Hamiltonian becomes even  larger if we 
write (\ref{e:seprime}) in the form that 
is fully analogous to the \schrod ~equation, for
we must rewrite it in terms
of normalized values.
Defining an inner product
as the integral over what we can call the
generalized waveguide's effective cross-section
(\textit{i.e.}, the region outside of which the EM wave's fields vanish, cf. Fig. \ref{f:picwg}),
one can 
introduce the normalization factor
\be\lb{e:wfnorm}
{\cal N}^2 
\equiv
%\left\langle \veeft | \veeft\right\rangle + \left\langle \vemft | \vemft\right\rangle \equiv
\int 
d
\textbf{x}_{\bot} 
\left(
\left|\veeft \right|^2  + \left| \vemft \right|^2
\right)
,
\ee
where $d \textbf{x}_{\bot}$ would be $d x d y$ in Cartesian coordinates.
Function ${\cal N}$ does not depend on the transverse coordinates but in general it can depend on $\t$.
With these definitions in hand,
we can define the following wavefunction (using the Dirac's bra-ket notations):
\be\lb{e:ketnorm}
\Psi
\equiv
\left\langle \textbf{x}_{\bot} | \Psi \right\rangle
=
\frac{1}{{\cal N}}
{\veeft\choose{\vemft}}
,
\ee
which is automatically normalized to one
\be
\left\langle \Psi | \Psi \right\rangle = 1
,
\ee
and thus the corresponding state vector $| \Psi \rangle$ can be regarded as a ray
in some appropriate Hilbert space (defined in Sec. \ref{s-ana-ev} below).
In terms of this state vector, Eq. (\ref{e:seprime}) 
acquires the Schr\"odinger equation's form
\be\lb{e:semain}
i \hbarr \pDer z 
\Psi 
=
\hamto
\Psi,
\ee
where
we denote the operator
%\bw
\ba
\hamto
&\equiv&
\hbarr 
\left(
\hamo
+
\hamto_{\cal N}
\right)
=
\hbarr 
\left(
\hat\sigma_2 \hamod
-
i
%\pDer z \!\left(\ln{{\cal N}}\right)
\gamman
\hat{\cal{I}}
\right)
\nn\\&=&
\hbarr
\twomat{0}{-\vee z \times \hat L_m}
{\vee z \times \hat L_e}{0}
-
i 
\hbarr
\gamman
%\pDer z \!\left(\ln{{\cal N}}\right)
\twomat{\hat I}{~0}{0}{~\hat I}\!,
~~~
\lb{e:hamto}
\ea
%\ew
where 
$\hat{\cal{I}}$ and $\hat I$ are, respectively,
the identity operator and the $2\times 2$ identity matrix,
and the coefficient
\be\lb{e:gamman}
\gamman  = \pDer z 
%\!\left(
\ln{|{\cal N}|}
%\right)
\ee
is in general a real-valued function of $\t$
(as well as a functional of the fields);
if ${\cal N}$ do vary with $z$ then Eq. (\ref{e:semain}) is not merely a rescaled version 
of Eq. (\ref{e:seprime}) but involves a corrective term, in general.
%$\hat{\cal{I}}$ and $\hat I$ are the unity operator and $2\times 2$ identity matrix, respectively,
Here by $z$ we assume the value $z/c$,
and the ``Planck'' constant $\hbarr$ 
is an effective scale constant of the dimensionality energy$\times$time,
which
is introduced for a purpose of preserving the correct dimensionality of the
relevant terms in the emergent Schr\"odinger equation
(the ambiguity of $\hbarr$ is yet another
manifestation of the absence of the fundamental length scale in Maxwell equations \cite{sybook09}).
Following traditions,
we will refer to the operator (\ref{e:hamto}) as the Hamiltonian operator of the system, although, strictly
speaking, a non-Hermitian operator cannot be fully regarded as Hamiltonian:
the anti-Hermitian part of such an operator does not correspond to
any symplectic structure
but rather plays a special role which will be discussed in Sec. \ref{s-nhgen-me} below.
It should be also noted that in a case when free charges or currents exist in the medium,
the original Maxwell equations of Sec. \ref{s-ana-ham}
must be modified, which can result either in a different
expression for our analogous Hamiltonian or in a breakdown of the Maxwell-Schr\"odinger analogy as such.
In this paper, we assume such charges and currents to be sufficiently small and consider
a leading-order approximation.

Furthermore,
it should be noticed the appearance 
of the additional term in the Hamiltonian (\ref{e:hamto}),
$
\hamto_{\cal N}
\equiv
-
i
%\pDer z \!\left(\ln{{\cal N}}\right)
\gamman
\hat{\cal{I}}
=
-
i 
%\pDer z \!\left(\ln{{\cal N}}\right)
\gamman
\twomatsm{\hat I}{~0}{0}{~\hat I}
,
$
which is essentially anti-Hermitian and proportional to the identity operator.
Due to the latter feature,
it belongs to the class of the Hamiltonian ``gauge'' terms, which role's
discussion will be postponed until Sec. \ref{s-nhgen-gau}.
 
Thus, equations (\ref{e:ketnorm})-(\ref{e:hamto}) 
represent the formal map between Maxwell equations for dielectric linear media
and the differential equation of the \schrod ~type,
which opens up the possibility of using quantum mechanical notions (with certain provisions of course)
for the purposes of a theory of EM wave propagation inside different dielectric materials, including waveguides.

%\sscn{Operator algebra and related observables}{s-ana-alg}

\sscn{Basic features of Hamiltonian}{s-ana-bas}

Let us consider now the operator (\ref{e:hamodef}).
Its Hermitian conjugate is given by
\be
\hamo^\dagger
=
\hamod^\dagger \hat\sigma_2
=
\twomat{0}{-\hat L_e^\dagger \vee z \times}
{\hat L_m^\dagger \vee z \times}{0}
,
\ee
where
\bse\lb{e:oplemhc}
\ba
&&
\hat L_e^\dagger =
\varepsilon^* \omega - \omega^{-1} \venat \times \frac{1}{\mu^{*}}\venat \times
,\lb{e:oplehc}\\&&
\hat L_m^\dagger =
\mu^* \omega - \omega^{-1} \venat \times \frac{1}{\varepsilon^*}\venat \times
,
\lb{e:oplmhc}
\ea
\ese
hence,
it is clear that $\hamo \not= \hamo^\dagger$ even if the $L$-operators
are both self-adjoint (which happens in the case $\mu = \mu^{*}$
and $\varepsilon = \varepsilon^*$, which will be discussed in what follows),
as $\hat{L}^\dagger$ now appears on the right side of the curl.
The only difference between the case of real-valued $\varepsilon$ and $\mu$
%(when both $\varepsilon$ and $\mu$ are real-valued) 
and the general one 
(when either or both $\varepsilon$ and $\mu$ are complex)
is that
the operator $\hamo$ is pseudo-Hermitian in the former case and strongly non-Hermitian
in the latter.
The eigenvalues of a pseudo-Hermitian operator are real-valued, whereas
the ones of a general non-Hermitian operator are complex.
Indeed, the case 
%when gain and loss are absent 
of real $\varepsilon$ and $\mu$
is a special one
because then
\be
\hamo^\dagger
=
\twomat{0}{-\hat L_e \vee z \times}
{\hat L_m \vee z \times}{0}
,
\ \
\forall
\ 
\varepsilon, \mu \in \Re
,
\ee
such that the following identity takes place
\be\lb{e:pseudoherm}
\hamo^\dagger \hat\sigma_2
-
\hat\sigma_2 \hamo
=
\hamod^\dagger
-
\hamod
=
0
,
\ \
\forall
\ 
\varepsilon, \mu \in \Re
,
\ee
which confirms that the operator $\hamo$ is pseudo-Hermitian
with $\hat\sigma_2$ playing a role of the intertwining operator \cite{zhu11}.
Alternatively, using (\ref{e:wfnorm}) and (\ref{e:ketnorm}),
one can check that the expectation value
of the operator $\hat\sigma_2 \hamo$
%$\left\langle \hat\sigma_2 \hamo \right\rangle$,
is real-valued when the values $\varepsilon$ and $\mu$ are --- 
since in that case the
operator $\hamod$ 
is self-adjoint, cf. Eq. (\ref{e:hamoddef}).

Besides, we can also derive that
\ba
&&
\hat\sigma_1 \hamo 
=
i
\twomat{\hat L_e}{0}
{0}{-\hat L_m}
,\\&&
(\hat\sigma_1 \hamo )^\dagger
\equiv
\hamo^\dagger \hat\sigma_1
=
-
\hat\sigma_1 \hamo
,
\ \
\forall
\ 
\varepsilon, \mu \in \Re
,\\
&&
\hat\sigma_3 \hamo 
=
-
\hamo  \hat\sigma_3 
=
-
\twomat{0}{\vee z \times \hat L_m}
{\vee z \times \hat L_e}{0}
%, \ \ \forall \ \varepsilon, \mu \in \Re
,
\ea
%from which one can derive the following relations:
such that 
\be
\left\{
\hat\sigma_3, \hamo 
\right\}
=
\left\{
\hat\sigma_3, \hat\sigma_2 \hamod 
\right\}
=
\hat\sigma_2
\left[
\hamod, \hat\sigma_3
\right]
=
0
,
\ee
and
\be
\left[
\hamod ,
\hat\sigma_3
\right]
=
\left[
\hamod^\dagger,
\hat\sigma_3
\right]
=
0
,
\ \
\forall
\ 
\varepsilon, \mu \in \Re
.
\ee
The last two equations become identities in the case of a diagonal matrix $\hamod$. 

For further developments it will be convenient to know the expectation values
of the products of operators that involve $\hamo$,
as well as their relations to the EM fields.
Using the formulas above, we obtain
\bw
\ba
\langle \hamo \rangle_\Psi
&=&
\frac{\omega}{{\cal N}^2}
\int d x d y 
\left[
\varepsilon
\vemft^* \cdot (\vee z \times \veeft)
-
\mu
\veeft^* \cdot (\vee z \times \vemft)
\right]
%\nn\\&&
-
\frac{i}{{\cal N}^2}
\int d x d y 
\left[
E_z
(\venat \cdot \veeft^*)
+
H_z
(\venat \cdot \vemft^*)
\right]
\nn\\
&=&
\frac{\omega}{{\cal N}^2}
\int d x d y 
\left(
\varepsilon
E_{[x} H_{y]}^*
+
\mu
E_{[x}^* H_{y]}
\right)
%\nn\\&&
-
\frac{i}{{\cal N}^2}
\int d x d y 
\left[
E_z
(\venat \cdot \veeft^*)
+
H_z
(\venat \cdot \vemft^*)
\right]
,\\%\ea \ba
\langle \hat\sigma_1 \hamo \rangle_\Psi
&=&
i
\frac{\omega}{{\cal N}^2}
\int d x d y 
\left(
\varepsilon
|\veeft|^2 
-
\mu
|\vemft|^2
+
\varepsilon^*
|E_z|^2 
-
\mu^*
|H_z|^2
\right)
,\ea\ba
\langle \hat\sigma_2 \hamo \rangle_\Psi
&=&
\langle \hamod \rangle_\Psi
=
\frac{\omega}{{\cal N}^2}
\int d x d y 
\left(
\varepsilon
|\veeft|^2 
+
\mu
|\vemft|^2
-
\varepsilon^*
|E_z|^2 
-
\mu^*
|H_z|^2
\right)
,\\
%\ea\ba
\langle \hat\sigma_3 \hamo \rangle_\Psi
&=&
\frac{\omega}{{\cal N}^2}
\int d x d y 
\left[
\varepsilon
\vemft^* \cdot (\vee z \times \veeft)
+
\mu
\veeft^* \cdot (\vee z \times \vemft)
\right]
%\nn\\&&
+
\frac{i}{{\cal N}^2}
\int d x d y 
\left[
E_z
(\venat \cdot \veeft^*)
-
H_z
(\venat \cdot \vemft^*)
\right]
\nn\\
&=&
\frac{\omega}{{\cal N}^2}
\int d x d y 
\left(
\varepsilon
E_{[x} H_{y]}^*
-
\mu
E_{[x}^* H_{y]}
\right)
%\nn\\&&
+
\frac{i}{{\cal N}^2}
\int d x d y 
\left[
E_z
(\venat \cdot \veeft^*)
-
H_z
(\venat \cdot \vemft^*)
\right]
,
\ea
and
\ba
\left\langle 
\pDer \omega 
\left(\hat\sigma_2 \hamo \right)
\right\rangle_\Psi
&=&
\left\langle 
\pDer \omega 
\hamod
\right\rangle_\Psi
=
\frac{1}{{\cal N}^2}
\int d x d y 
\left(
\varepsilon
|\veeft|^2 
+
\mu
|\vemft|^2
+
\varepsilon^*
|E_z|^2 
+
\mu^*
|H_z|^2
\right)
,
\ea
and similarly for the adjoint operator:
\ba
\langle \hamo^\dagger \rangle_\Psi
&=&
\frac{\omega}{{\cal N}^2}
\int d x d y 
\left[
\mu^*
\vemft^* \cdot (\vee z \times \veeft)
-
\varepsilon^*
\veeft^* \cdot (\vee z \times \vemft)
\right]
%\nn\\&&
+
\frac{i}{{\cal N}^2}
\int d x d y 
\left[
E_z^*
(\venat \cdot \veeft)
+
H_z^*
(\venat \cdot \vemft)
\right]
\nn\\
&=&
\frac{\omega}{{\cal N}^2}
\int d x d y 
\left(
\varepsilon^*
E^*_{[x} H_{y]}
+
\mu^*
E_{[x} H^*_{y]}
\right)
%\nn\\&&
+
\frac{i}{{\cal N}^2}
\int d x d y 
\left[
E_z^*
(\venat \cdot \veeft)
+
H_z^*
(\venat \cdot \vemft)
\right]
,\\
\langle \hat\sigma_1 \hamo^\dagger \rangle_\Psi
&=&
i
\frac{\omega}{{\cal N}^2}
\int d x d y 
\left(
\mu^*
|\veeft|^2 
-
\varepsilon^*
|\vemft|^2
\right)
%\nn\\&&
-
\frac{i}{{\omega \cal N}^2}
\int d x d y 
\left(
\frac{1}{\varepsilon^*}
\left|
\venat \cdot \veeft
\right|^2
-
\frac{1}{\mu^*}
\left|
\venat \cdot \vemft
\right|^2
\right)
,\\
\langle \hat\sigma_2 \hamo^\dagger \rangle_\Psi
&=&
\frac{\omega}{{\cal N}^2}
\int d x d y 
\left(
\mu^*
|\veeft|^2 
+
\varepsilon^*
|\vemft|^2
\right)
%\nn\\&&
-
\frac{1}{{\omega \cal N}^2}
\int d x d y 
\left(
\frac{1}{\varepsilon^*}
\left|
\venat \cdot \veeft
\right|^2
+
\frac{1}{\mu^*}
\left|
\venat \cdot \vemft
\right|^2
\right)
,\\
\langle \hat\sigma_3 \hamo^\dagger \rangle_\Psi
&=&
-
\frac{\omega}{{\cal N}^2}
\int d x d y 
\left[
\mu^*
\vemft^* \cdot (\vee z \times \veeft)
+
\varepsilon^*
\veeft^* \cdot (\vee z \times \vemft)
\right]
%\nn\\&&
-
\frac{i}{{\cal N}^2}
\int d x d y 
\left[
E_z^*
(\venat \cdot \veeft)
-
H_z^*
(\venat \cdot \vemft)
\right]
\nn\\
&=&
\frac{\omega}{{\cal N}^2}
\int d x d y 
\left(
\varepsilon^*
E^*_{[x} H_{y]}
-
\mu^*
E_{[x} H_{y]}^*
\right)
%\nn\\&&
-
\frac{i}{{\cal N}^2}
\int d x d y 
\left[
E_z^*
(\venat \cdot \veeft)
-
H_z^*
(\venat \cdot \vemft)
\right]
.
\ea
\ew

It is easy to see that some of these expectation values
can be related to the energy density quantities 
of a wave mode
\ba
&&
\av{\hat\sigma_2 \hamo}_\Psi
=
\langle 
\hamod \rangle_\Psi
=
\frac{
%16 \pi 
4
\omega}{{\cal N}^2}
\left(
{\cal E}_\bot
-
{\cal E}_z^*
\right)
,\lb{e:avener1}\\&&
\langle 
{\cal L}_+
(
\hat\sigma_2 \hamo
)
\rangle_\Psi
= 
\langle
{\cal L}_+
\hamod
\rangle_\Psi
%\nn\\&&\qquad\qquad
= 
\frac{
%32 \pi 
8
\omega}{{\cal N}^2}
{\cal E}_\bot
,\\&&
\langle 
{\cal L}_-
(
\hat\sigma_2 \hamo
)
\rangle_\Psi^*
= 
\langle
{\cal L}_-
\hamod
\rangle_\Psi^*
%\nn\\&&\qquad\qquad
= 
\frac{
%32 \pi 
8
\omega}{{\cal N}^2}
{\cal E}_z
,\\&&
\langle
{\cal L}_+
\hamod
\rangle_\Psi
+
\langle
{\cal L}_-
\hamod
\rangle_\Psi^*
%\nn\\&&\qquad\qquad
= 
\frac{
%32 \pi 
8
\omega}{{\cal N}^2}
{\cal E}_\text{tot}
,
\lb{e:avener4}
\ea
where
${\cal L}_\pm = \omega \pDer \omega \pm 1$ is a differential operator with respect to 
the frequency parameter,
and
\bse
\ba
&&
{\cal E}_z
=
\frac{1}{
%16 \pi
4}
\int d x d y 
\left(
\varepsilon
|E_z|^2 
+
\mu
|H_z|^2
\right)
,\\
&&
{\cal E}_\bot
=
\frac{1}{
%16 \pi
4}
\int d x d y 
\left(
\varepsilon
|\veeft|^2 
+
\mu
|\vemft|^2
\right)
,
\ea
\ese
are, respectively, the longitudinal and transverse 
%time-averaged 
EM wave energy
densities,
such that
\be\lb{e:avenertot}
{\cal E}_\text{tot} = {\cal E}_\bot + {\cal E}_z
\ee
is the total energy density of a mode described by the state vector
(\ref{e:ketnorm}).
Finally, one can add to this list the transmitted power 
$\tpower$
from Eq. (\ref{e:onemodetpower}).

\sscn{Hilbert space and $L$-diagonal representation}{s-ana-ev}

Let us consider the following auxiliary eigenvalue problem:
\bse\lb{e:evoplem}
\ba
&&
\hat L_e \eigf e = \eigv{e} \eigf e
,\lb{e:evople}\\&&
\hat L_m \eigf m = \eigv{m} \eigf m
,
\lb{e:evoplm}
\ea
\ese
where the eigenfunctions $\eigf{e),(m}$ are 2D vectors
obeying the normalization condition
\be
\int d x d y 
\left(
|\eigf e|^2 
+
|\eigf m|^2
\right)
=
1
,
\ee
as well as the boundary conditions -- they must
%for this problem is that the eigenfunctions must
vanish either outside the medium or
at the transverse spatial infinity 
%$\left\{|x|, |y| \right\} \to +\infty $
$|x|^2 + |y|^2 \to +\infty $.
The auxiliary eigenvalues 
%$\eigv{}$'s 
are in general complex-valued functions of $\t$ and $\omega$,
$\eigv{} = \eigv{}(\t,\omega)$,
which also depend on the parameters inside the functions
$\varepsilon (x,y,z)$ and $\mu (x, y, z)$.
They can become real-valued in some cases, e.g., when the permittivity and permeability
are both real-valued. 
In the latter case, both $L$-operators become self-adjoint, therefore, this eigenvalue problem 
reduces to that of the Sturm-Liouville type.

Thus, 
%one can see that 
the total Hilbert space of our theory can be represented
as
a direct sum of the function space of the eigenvectors $\eigf{e),(m}$.
These spaces store, separately from each other, all electric and magnetic
components of EM wave modes, which are allowed by imposed
boundary conditions.
Their union space thus contains  the information about energy transfer between
electric and magnetic components for each mode, see Appendix \ref{a:op} for details.
Below in Sec. \ref{s-tls-rea}, we will see
%, for instance when we will use a plane-wave ansatz 
that
the auxiliary eigenvalues $\eigv{(e),(m)}$ have distinct behaviors compared to the
familiar eigenvalues of Helmholtz-type equation
% (wavevectors, k(\omega))
and their dispersion relations
%$\lambda(\omega)$ 
show the physics of
waves under a different light.

Further,
in this representation, the operators (\ref{e:hamodef})
and (\ref{e:hamoddef})
take the form
\ba
\hamod
&= &
\twomat{\eigv{e} \hat I}{0}
{0}{\eigv{m} \hat I}
=
\eigv{-}
\hat\sigma_3
+
\eigv{+}
\hat{\cal I}
,
\lb{e:hamoddefII}
\\
\hamo
&= &
\hat\sigma_2 \hamod
=
i \eigv{-}
\hat\sigma_1
+
\eigv{+}
\hat\sigma_2
,
\lb{e:hamodefII}
\ea
where 
\be
\eigv{\pm} = \tfrac{1}{2} (\eigv{e} \pm \eigv{m})
,
\ee
therefore, the total Hamiltonian (\ref{e:hamto}) becomes
%\bw
\ba
\hamto
%= \hamo + \hamto_{\cal N}
&=&
\hbarr
\hat\sigma_2
\twomat{\eigv{e} \hat I}{0}
{0}{\eigv{m} \hat I}
-
i 
\hbarr
\gamman
\twomat{\hat I}{~0}{0}{~\hat I}
\nn\\
&=&
\hbarr
\left(
i \eigv{-}
\hat\sigma_1
+
\eigv{+}
\hat\sigma_2
-
i
\gamman
\hat{\cal{I}}
\right)
\nn\\
&=&
i
\hbarr
\left(
\eigv{e}
\hat\sigma_-
-\eigv{m}
\hat\sigma_+
-
\gamman
\hat{\cal{I}}
\right)
,
\lb{e:hamtoII}
\ea
where $\hat\sigma_{\pm} = \tfrac{1}{2} (\hat\sigma_1 \pm i \hat\sigma_2)$.
Thus, in this representation the Hamiltonian becomes 
a 4$\times$4 matrix consisting of 2$\times$2 blocks.

\scn{Statistical mechanics of wave modes}{s-nhgen}

What we have done in the previous section is merely a way of rewriting Maxwell equations
for waves in dielectric media
in the Schr\"odinger 
form (\ref{e:semain}).
In this section we will proceed with an important generalization:
we go beyond those equations and
adapt the quantum-type density matrix approach \cite{fbook,sz13,sz14,sz14cor,ser15,sz15,zlo15}
for describing
the propagation of EM waves 
inside dielectric media.
% described by non-Hermitian Hamiltonian operators.
This will allow us to describe not only separate wave modes (``pure states'', in quantum-mechanical terminology)
or
their superpositions (``entangled pure states'')
but also their statistical ensembles (``mixed states'').
The latter are crucial for introducing the dissipative effects since the purity
of the states is not necessarily preserved in presence of dissipative environments \cite{zur96,zlo15}.
More details will be given in Sec. \ref{s-nhgen-me} below, here we only note that
one should not confuse our approach with the Loudon 
microscopical QED-type quantization of light in presence of matter
\cite{lobook,jil93}, nor our approach is directly related
to the Nyquist-Callen(-Rytov) approach to thermal fluctuational electrodynamics \cite{ryt58}.
The quantum-type statistics of wave modes we are going to introduce here is an effective phenomenon 
which emerges due to
existence of the Maxwell-Schr\"odinger map and associated quantum-type dynamics.

Furthermore, the main difference of the proposed approach from the standard non-Hermitian 
quantum-statistical one \cite{fbook,sz13,sz14,sz14cor,ser15,sz15,zlo15}
is that the role of the time variable is played here by the third coordinate, $z/c$.
In other words, instead of time evolution of quantum states the method will describe
the distribution of EM wave energy along the propagation axis.
This, however, does not pose
much difference from the technical viewpoint, and most of concepts can
be borrowed and applied for the purposes of the EM theory.

Finally, due to the fact that in
this theory
both the speed of light and effective Planck constant 
are scale constants, 
from now on
we work in units where $\hbarr = \hbar = c = 1$.

\sscn{Master equation}{s-nhgen-me}

To begin with, if a Hamiltonian is a non-Hermitian operator, then
it can be decomposed into its Hermitian and anti-Hermitian parts, respectively:
\be\lb{e:nhham}
\hamto  = \hamto_+ + \hamto_-
=
\hamto_+  - i \hat\Gamma
,
\ee
where we use the notations
\be
\hamto_\pm 
\equiv
\frac{1}{2}
\left(
\hamto \pm \hamto^\dagger
\right)
=
\pm \hamto_\pm^{\dag}
,
\ee
and the Hermitian operator
\be
\hat\Gamma \equiv i \hamto_- 
=
\hat\Gamma^{\dag} 
\ee
is usually dubbed the \textit{decay operator}.
For instance, for the Hamiltonian (\ref{e:hamto}) 
one easily computes that
\bse\lb{e:hamtoHaH}
\ba
&&
\hamto_+
=
{\hamo}_+
=
\frac{1}{2}
\left(
\hat\sigma_2 \hamod + \hamod^\dagger \hat\sigma_2
\right)
,\lb{e:hamtoH}\\&&
\hat\Gamma
=
\decoO
+
\gamman
\hat{\cal{I}}
=
\frac{i}{2}
\left(
\hat\sigma_2 \hamod - \hamod^\dagger \hat\sigma_2
\right)
+
%\pDer z \!\left(\ln{{\cal N}}\right)
\gamman
\hat{\cal{I}}
,
\lb{e:hamtoAH}
\ea
\ese
where we assume the notations from the previous section.
This decomposition means that within the total system, described by $\hamto$, one can
single out the Hermitian subsystem, described by $\hamto_+$,
whereas the operator $\hat\Gamma$ can be regarded as describing the energy
exchange of this subsystem with its environment.
The question of whether the system $\hamto$ itself can be a subsystem of a more general, Hermitian, system, is not considered here, since it would bring us outside the scope of this paper.

The quantum-statistical approach means here that
the ``evolution'' (distribution along the propagation direction) of such a system  is described
by the (reduced) density operator,
%$\hat\rho = \hat\rho (z)$, 
which 
%is Hermitian, positive semi-definite, and has a unit trace. This operator
contains information not only about superpositions
of the EM wave modes
but also about
the statistical uncertainty
of their distribution inside a medium.
%which is not related to Maxwell equations for EM waves.
Such uncertainty can be caused, for instance, by the interaction
of the wave with its environment,
which usually happens inside realistic dielectric media.
An example would be the thermal randomness that arises
in the statistical mixture of large numbers of EM wave modes, 
each with a certain classical probability, switching from one to another due to thermal fluctuations. 
In such cases, unpolarized light (``mixed state'') appears, 
which is in fact
not the plain superposition of single modes (``pure states''), but 
their statistical ensemble.
Thus,
the density matrix contains all the information necessary to calculate 
any measurable property of polarized or unpolarized radiation
propagating inside realistic media with or without dissipation.
Besides,
one of its advantages 
is that for each mixed state
there can be many statistical ensembles of pure states
but
only one density matrix.

In our case, a density operator has a few distinctive features when it comes to
its interpretation.
Firstly, it is a reduced density operator which means that environment's degrees of freedom
have been averaged out, one deals only with their cumulative effect upon the subsystem
described by such a density operator.
Furthermore, the Hilbert space of the Hamiltonian (\ref{e:hamto}) 
has a block structure where one block corresponds to a set of all electric 
components of EM wave modes and other block does to all magnetic ones, see Sec. \ref{s-ana-ev}.
%Therefore, from quantum viewpoint, this Hamiltonian describes the ``intrinsic'' energy transfer between ``bands'' representing electric and magnetic components of EM wave modes which are allowed by imposed boundary conditions; it is obvious that this process must be affected by environment.
Thus, the off-diagonal $m n$th components of the density matrix describe the transition between
the electrical component of $m$th mode and magnetic component of $n$th mode, 
%(or \textit{vice versa}), 
where indexes belong to different blocks,
whereas the diagonal components are related to integral energy measures of either
electrical or magnetic components of a single mode, see Sec. \ref{s-nhgen-ave}, 
Eqs. (\ref{e:decompom}), (\ref{e:decomprho}),  and Appendix \ref{a:op}.

An equation for the density matrix can be directly derived from any equation that
has the \schrod ~form, 
see, for instance, Ref. \cite{fbook}.
Using Eq. (\ref{e:semain}), one can show that our NH system is fully described
by the so-called non-normalized
density operator $\densnnorm$, which is defined as a solution
of the operator equation of the Liouvillian type,
\be\label{e:densnonnorm}
\frac{d}{d \t}
%\partial_t
\densnnorm
=
i 
\left(
\densnnorm \hamto^\dagger
-
\hamto \densnnorm
\right)
=
- i \left[\hamto_+,\densnnorm\right]
-
\left\{\hat{\Gamma},\densnnorm\right\}
,
\ee
where 
%we denoted \be \t = z/c ,\ee and the 
square and curly brackets 
%on the right hand side of Eq. (\ref{e:densnonnorm})
denote the commutator and anti-commutator, respectively.
One can see that, as $z$ varies, the trace of $\densnnorm$ is not conserved,
\begin{equation}
\frac{d}{d \t}
{\rm tr}\,\densnnorm
=
-
%\frac{2}{\hbar}
2
\avo{\hat{\Gamma}}
,
\label{e:dotTrOmega}
\end{equation}
where we denoted
\be
\avo{\hat{\Gamma}}
=
{\rm tr}
(\hat{\Gamma} \,\densnnorm
)
,
\ee
therefore,
$\densnnorm$ describes a case when 
%a subsystem is known to certainly disappear:
subsystem's integrity will eventually be 
completely broken -- either through the complete decay (${\rm tr}\densnnorm$ vanishes at large $\t$)
or critical instability (${\rm tr}\densnnorm$ blows up at large $\t$).
In both cases, the subsystem 
gets destroyed very fast, usually at an exponential rate.

This is definitely not what always happens in reality,
therefore
%in order to define a density operator for a sustainable system (the physical meaning of this will be discussed later),
in Ref. \cite{sz13}  we introduced
the operator
\be\lb{e:rhonorm}
\hat\rho = \densnnorm / {\rm tr}\, \densnnorm
,
\ee
which is automatically normalized (the physical meaning of this procedure will be discussed later),
therefore, it can be used for computing
expectation values, correlation functions and other observables.

In principle, in Eq. (\ref{e:densnonnorm}) one can change from $\densnnorm$ to $\hat\rho$, 
and obtain the equation for
the normalized density operator itself
\begin{eqnarray}
\frac{d}{d \t}
%\partial_t
\hat\rho
=
- i
\left[\hamto_+,\hat\rho
\right]
-
\left\{\hat{\Gamma},\hat\rho 
\right\}
+
2
\langle \Gamma \rangle
\hat\rho
,
\label{e:eomrho}
\end{eqnarray}
where 
the notation 
\be\label{e:defmean}
\av A = {\rm tr} (\hat\rho \, \hat{A})
\ee
will be used 
for denoting the expectation value of any given operator $\hat{A}$
with respect to the normalized density operator.
%which determines $\hat\rho = \hat\rho (t)$ directly but becomes nonlinear.
%It should be noted, however, that Eq. (\ref{e:eomrho}) contains slightly less information about the system (\ref{e:nhham}) than (\ref{e:densnonnorm}) because the procedure (\ref{e:rhonorm}) erases the information about the overall factor of $\densnnorm$ including its trace. This missing piece of information can be useful, e.g.,  when studying the initial conditions or entropic properties of the system, see, respectively, Secs. \ref{s-nhgen-ini} and \ref{s-nhgen-entr} for details.

From the mathematical point of view, 
Eq. (\ref{e:eomrho}) is both nonlocal and nonlinear
with respect to the density operator $\hat\rho$.
Though, this does not pose a significant problem from the technical point
of view, since one can always use Eq. (\ref{e:rhonorm})
as an ansatz for getting a linear equation. 
Thus, Eqs. (\ref{e:densnonnorm})-(\ref{e:eomrho}), 
together with the definition for computing the expectation values (\ref{e:defmean}),
represent the map that allows us to describe the distribution of  
system (\ref{e:nhham}) along $z$ direction
in terms of the matrix differential equation, whose
%can be  defined on the mathematical foundations that are 
mathematical structure
resembles the one of the conventional 
master equations of the Lindblad kind  \cite{lin76}.
According to this map, the Hermitian operator 
$\hamto_+ =  (\hamto + \hamto^\dagger)/2$ takes
over a role of the system's Hamiltonian (cf. the commutator term in equations (\ref{e:densnonnorm}) 
or (\ref{e:eomrho}) above)
whereas
the decay operator $\hat\Gamma = i (\hamto - \hamto^\dagger)/2 $ induces
additional terms in the evolution equation
that are supposed to account for NH effects.
In other words, a theory with the non-Hermitian Hamiltonian $\hamto$ is dual to
a theory with the Hermitian Hamiltonian $(\hamto + \hamto^\dagger)/2$
but with the modified evolution equation, which thus becomes the master equation of a special
kind.
This equivalence not only 
reveals new features of the dynamics driven by non-Hermitian Hamiltonians 
but also facilitates the application of such Hamiltonians
for open quantum systems \cite{sz14}.

From the viewpoint of theory of open quantum systems,
the equation
for the non-normalized density operator $\densnnorm$
effectively describes the subsystem represented by the Hamiltonian $\hamto_+$ with the effect of 
environment represented by $\hat{\Gamma}$.
If the ``evolution'' (energy distribution along $\t$ direction) is governed by $\densnnorm$,
which trace is not preserved,
then this subsystem ``eventually'' (at some value of $\t$) becomes critically unstable or disappears.
Thus, the post-selecting  procedure (\ref{e:rhonorm}) can be interpreted
as follows:
in order to maintain the probabilistic interpretation of $\hat\rho$
(as well as to ensure that the subsystem exists at every point $z$),
one must neglect the amount
of the energy that is not associated with
the original subsystem $\hamto_+$ anymore.
Consequently,
the equation
for the normalized density operator $\hat\rho$
effectively describes the subsystem $\hamto_+$ together with the effect of 
environment  $\hat{\Gamma}$ 
and the energy flow between this subsystem and environment.

To summarize, the normalized $\hat\rho$ and non-normalized $\densnnorm$ density operators
describe, at a given Hamiltonian and initial and boundary conditions, two types of EM wave evolution.
The former operator
applies
if the wave's evolution is knowingly \textit{sustainable};
%\textit{i.e.}, the system's probability space does not shrink or expand, 
in this case the operator $\densnnorm$ 
%should not be discarded since it 
contains the information about the above-mentioned  energy flow
between the system and environment, which can be extracted
using auxiliary techniques (such as the entropy, see Sec. \ref{s-nhgen-entr}
below).
If the normalized density operator does not exist or it has unphysical properties
(\textit{e.g.}, singularity at some value of $\t$)
then one is left with the \textit{non-sustainable} evolution described solely by the operator $\densnnorm$.
The choice between these types of evolution is dictated by the physical context:
it depends on values of Hamiltonian's parameters as well as on boundary and initial conditions for master equations, which specify one or another 
configuration.

\sscn{Initial conditions}{s-nhgen-ini}

%\textit{Initial/boundary conditions}.

Obviously, 
any Liouvillian-type dynamics implies that the equations for density operators must be supplemented with  
initial conditions.
In our case, a few subtle points exist that must be taken into account.
First, since the role of time is played by the $\t$-coordinate here, the initial condition
at the surface $\t = \t_0$ is, strictly speaking, a boundary condition.
For example, it is convenient to choose this surface to be the interface between a medium
and the rest of space,  which is
orthogonal to the propagation direction.
Thus, the choice must be dictated by geometrical properties
of the material layout.

The second subtlety is that, in our case, we have two possible
types of evolution, described by two equations --
for the non-normalized $\densnnorm$ and normalized $\hat\rho$
density operators --
equations (\ref{e:densnonnorm}) and (\ref{e:eomrho}), respectively.
%While the operator $\hat\rho$ is more suitable for statistical description, it is the operator $\densnnorm$ which is the ``primordial'' object of the theory.
However, it is Eq. (\ref{e:densnonnorm}), which must
be solved in the first place in both cases, therefore
the corresponding initial/boundary condition at the surface $\t = \t_0$
must be specified as
\be\lb{e:doinigen}
%\hat\rho (\t =0) 
%|_{\t = 0} = 
\densnnorm
%(\t = \t_0) 
|_{\t = \t_0}
= \densnnorm_0
,
\ee
whereas for the normalized density operator we would automatically obtain
\be\lb{e:ndoinigen}
%\hat\rho (\t =0) 
%|_{\t = 0} = 
\hat\rho 
%(\t = \t_0) 
|_{\t = \t_0}
= \densnnorm_0 / \text{tr}\, \densnnorm_0
,
\ee
according to Eq. (\ref{e:rhonorm}).
Note that from a physical point of view,
the condition (\ref{e:doinigen}) is not equivalent to
$\hat\rho (\t = \t_0)  = \densnnorm_0$ since the operator $\densnnorm_0$
does not necessarily have a unit trace,
hence fixing its trace's
%the $\text{tr} \densnnorm_0$'s 
value would
require invocation of additional physical considerations.

The third subtlety arises when dealing with pure states, \textit{i.e.},
states whose density matrices $\hat\varrho$ obey the idempotency condition
${\hat\varrho}^2 = \hat\varrho$.
When dealing with Hermitian Hamiltonians,
it is often convenient to choose a pure state as an initial one because
pure states have simpler structure and are easier to prepare.
However, in our case, the original incident wave is not necessarily in a pure
state, therefore, not all initial/boundary conditions are admissible.
Besides, even if the operator $\densnnorm_0$ is pure
then it does not necessarily mean that the
operator $\densnnorm_0 / \text{tr}\, \densnnorm_0$ will also be pure.

To summarize, when dealing with EM wave's propagation in media
within the framework of the statistical approach,
there is no ``conventional'' set of values of $\densnnorm_0$,
instead these must be decided on a case by case basis, depending on the physical context.
Besides, since the Hamiltonian (\ref{e:hamto})
depends not only on 
permittivity and permeability but also on the values of EM fields, the  properties of the whole system depend both
on the properties of a medium and on the characteristics of the original (incident)
wave. 
Therefore, one must always refer to the total  ``medium+wave'' configuration
when describing properties of our system.

\sscn{Averages and observables}{s-nhgen-ave}

The simplest averages one can start from are the primary
ones, by which we imply the expectation values of the Pauli operators,
\ba
&&
\av{\sigma_a}
=
\text{tr}
\left(
\hat\rho \hat\sigma_a
\right)
=
\avo{ 
\sigma_a
}/
\text{tr}
\densnnorm
,
\lb{e:primav}
\\&&
\avo{\sigma_a}
=
\text{tr}
(
\densnnorm \hat\sigma_a
)
,
\ea
where $a =1,2,3$;
here $\av{\sigma_a}$ are 
the expectation values for sustainable evolution (see Sec. \ref{s-nhgen-me}), whereas
$\avo{\sigma_a}$ 
are
the ones for non-sustainable evolution.
In the theory of EM wave propagation,
%the physical meaning of 
these averages
%$\left\langle \sigma_i \right\rangle$ 
would be related to the energy flow between the wave
modes and a dielectric medium, see Appendix \ref{a:op}.
The primary averages are also useful when one needs
to decompose a given density operator in terms of the Pauli 
operators.\\

%\bc
\textit{(a) Main equations}. Using Eqs. (\ref{e:hamtoHaH})-(\ref{e:defmean}),
one easily obtains the equations for the averages
\bse
\ba
\frac{d}{d \t}
\avo{\sigma_1}
&=&
\avo{\hat\sigma_3 \hamod + \hamod^\dagger \hat\sigma_3}
-
2 \Gamma_{\cal N} \avo{\sigma_1}
,
\\
\frac{d}{d \t}
\avo{\sigma_2}
&=&
i
\avo{\hamod^\dagger - \hamod}
-
2 \Gamma_{\cal N} \avo{\sigma_2}
,
\\
\frac{d}{d \t}
\avo{\sigma_3}
&=&
-
\avo{\hat\sigma_1 \hamod + \hamod^\dagger \hat\sigma_1}
-
2 \Gamma_{\cal N} \avo{\sigma_3}
,
\\
\frac{d}{d \t}
\text{tr}\, \densnnorm
&=&
%- 2 \avo{\Gamma}=
i
\avo{
\hamod^\dagger \hat\sigma_2
-
\hat\sigma_2 \hamod
}
-
2 \Gamma_{\cal N}\, \text{tr} \densnnorm
,
\ea
\ese
and
\bse\lb{e:primaveq}
\ba
\frac{d}{d \t}
\av{\sigma_1}
&=&
\av{\hat\sigma_3 \hamod + \hamod^\dagger \hat\sigma_3}
+
i
\av{
\hat\sigma_2 \hamod
-
\hamod^\dagger \hat\sigma_2
}
\av{\sigma_1}
,
\\
\frac{d}{d \t}
\av{\sigma_2}
&=&
i
\av{\hamod^\dagger - \hamod}
+
i
\av{
\hat\sigma_2 \hamod
-
\hamod^\dagger \hat\sigma_2
}
\av{\sigma_2}
,
\lb{e:primaveq2}\\
\frac{d}{d \t}
\av{\sigma_3}
&=&
-
\av{\hat\sigma_1 \hamod + \hamod^\dagger \hat\sigma_1}
+
i
\av{
\hat\sigma_2 \hamod
-
\hamod^\dagger \hat\sigma_2
} 
\av{\sigma_3}
,
~~~~
\ea
\ese
where 
we have used the notations
$\avo{\hat A} = \text{tr} (\densnnorm \hat A)$.
For instance, in the $L$-representation (see Sec. \ref{s-ana-ev}),
in matrix notations
these equations become, respectively,
\be\lb{e:primavoLeqL}
%\begin{matrix}
\frac{1}{2}
\frac{d}{d \t}
\fourcol{
\avo{\sigma_1}}{
\avo{\sigma_2}}{
\avo{\sigma_3}}{
\text{tr} \densnnorm
}
=
\begin{pmatrix}
- \Gamma_{\cal N} & 0 & \evre_+ & \evre_- \\[0.3ex]
0 & - \Gamma_{\cal N} & - \evim_- & - \evim_+ \\[0.3ex]
- \evre_+ &  \evim_- & - \Gamma_{\cal N} & 0  \\[0.3ex] 
\evre_- &  - \evim_+ & 0 & - \Gamma_{\cal N} 
\end{pmatrix}
\!
\fourcol{
\avo{\sigma_1}}{
\avo{\sigma_2}}{
\avo{\sigma_3}}{
\text{tr} \densnnorm
}
,
\ee
and
\be\lb{e:primaveqL}
%\begin{matrix}
\frac{1}{2}
\frac{d}{d \t}
\threecol{
\av{\sigma_1}}{
\av{\sigma_2}}{
\av{\sigma_3}}
=
\begin{pmatrix}
\av{\decoO} & 0 & \evre_+ \\
0 & \av{\decoO} & - \evim_- \\
- \evre_+ &  \evim_- &  \av{\decoO} 
\end{pmatrix}
\!
\threecol{
\av{\sigma_1}}{
\av{\sigma_2}}{
\av{\sigma_3}}
+
\threecol{\evre_-}{-\evim_+}{0}
,
\ee
where 
\be\lb{e:avGammapL}
\av{\decoO}
\equiv
\frac{i}{2}
\av{\hamo - \hamo^\dagger}
=
- \evre_- \av{\sigma_1} + \evim_- \av{\sigma_2}
\ee
and the matrix components
\bse\lb{e:coeffRI}
\ba
\evre_\pm &=&
\frac{1}{2} (\eigv{\pm} + \eigv{\pm}^*) = 
\frac{1}{4} \left[\eigv{e} +  \eigv{e}^* \pm (\eigv{m} + \eigv{m}^*)\right]
,~~~~\lb{e:coeffR}\\
\evim_\pm &=&
\frac{i}{2} (\eigv{\pm} - \eigv{\pm}^*) = 
\frac{i}{4} \left[\eigv{e} -  \eigv{e}^* \pm (\eigv{m} - \eigv{m}^*)\right]
,
\lb{e:coeffI}
\ea
\ese
are real-valued numbers;
the last two equations can be
inverted and written in the  form:
\bse\lb{e:coeffReeImm}
\ba
&&
\text{Re} (\eigv{e}) 
\equiv 
\frac{1}{2}
\left(
\eigv{e} + \eigv{e}^*
\right)
=
\evre_+ + \evre_-
,\lb{e:coeffRee}\\&&
\text{Im} (\eigv{e}) 
\equiv 
\frac{1}{2 i}
\left(
\eigv{e} - \eigv{e}^*
\right)
=
- \evim_+ - \evim_-
,\\&&
\text{Re} (\eigv{m}) 
\equiv 
\frac{1}{2}
\left(
\eigv{m} + \eigv{m}^*
\right)
=
\evre_+ - \evre_-
,\\&&
\text{Im} (\eigv{m}) 
\equiv 
\frac{1}{2 i}
\left(
\eigv{m} - \eigv{m}^*
\right)
=
- \evim_+ + \evim_-
,\lb{e:coeffImm}
\ea
\ese
which will be also useful in what follows.

Correspondingly, the
``steady-state'' (extremum) values of 
the primary averages for sustainable evolution can be found as a solution
the quadratic equations
\be\lb{e:primavLstd}
%\begin{matrix}
\begin{pmatrix}
\av{\decoO}_s & 0 & \evre_+ \\
0 & \av{\decoO}_s & - \evim_- \\
- \evre_+ &  \evim_- &  \av{\decoO}_s 
\end{pmatrix}
\!
\threecol{
\av{\sigma_1}_s}{
\av{\sigma_2}_s}{
\av{\sigma_3}_s}
=
\threecol{-\evre_-}{\evim_+}{0}
,
\ee
where
\be\lb{e:avGammapLstd}
\av{\decoO}_s
=
- \evre_- \av{\sigma_1}_s + \evim_- \av{\sigma_2}_s
\ee
is a ``steady-state'' (extremum) value of the average of the 
$\decoO$
operator.\\

%\bc
\textit{(b) Energy density}.
%\ec
By analogy with Eqs. (\ref{e:avener1})-(\ref{e:avenertot}),
the wave's energy density averaged over
the statistical
ensemble represented by $\hat\rho$
can be written as
\ba
&&
\wenbar_\bot
=
\frac{{\cal N}^2}{
%32 \pi 
8\omega}
\av{
{\cal L}_+
\hamod
}
%\nn\\&&\qquad\qquad
,\\&&
\wenbar_z
=
\frac{{\cal N}^2}{
%32 \pi 
8
\omega}
\av{
{\cal L}_-
\hamod
}^*
%\nn\\&&\qquad\qquad
,\\&&
\wenbar_\text{tot} = \wenbar_\bot + \wenbar_z
,
\lb{e:avenertotens}
\ea
where the differential operators ${\cal L}_\pm = \omega \pDer \omega \pm 1$
were already defined after Eq. (\ref{e:avener4}),
and
$\wenbar_z$, $\wenbar_\bot$ and $\wenbar_\text{tot}$
are, respectively, longitudinal, transverse and total energy
densities.
Thus, 
%from the viewpoint of the classical electrodynamics, 
the ratio
\be
\Xi_\wen
=
\frac{
\wenbar_z}{\wenbar_\bot}
=
\frac{
\av{
{\cal L}_-
\hamod
}^*
}{
\av{
{\cal L}_+
\hamod
}}
\ee
describes how much of the averaged wave energy is stored in the longitudinal component
as compared to the transverse one.
Other related ratios can be expressed via $\Xi_\wen$: 
\be
\Xi_\bot
\equiv
\frac{\wenbar_\bot}{ \wenbar_\text{tot} }
= 
1 - \frac{\wenbar_z}{\wenbar_\text{tot} }
=
\frac{1}{1+\Xi_\wen}
,
\ee
and so on.

Besides, using the corresponding formulas from Appendix \ref{a:op},
one can add to this list the following energy-related identities
%\bse\lb{e:vwenbarem}
\ba
&&
\vwenbar_\bot^{(e)}
=
{\cal N}^2
\av{\hat e}
=
\frac{{\cal N}^2}{2} 
\left(
1
+
\av{
\hat\sigma_3
}
\right)
,\lb{e:vwenbare}\\&&
\vwenbar_\bot^{(m)}
=
{\cal N}^2
\av{\hat g}
=
\frac{{\cal N}^2}{2} 
\left(
1
-
\av{
\hat\sigma_3
}
\right)
,
\lb{e:vwenbarm}
\ea
%\ese
where
$\vwenbar_\bot^{(e)}$ and $\vwenbar_\bot^{(m)}$
are, respectively, the electric and magnetic components of the transverse energy
density averaged over the ensemble $\hat\rho$, 
in absence of the effect of the medium: $\vwenbar = \wenbar|_{\varepsilon, \mu \to 1}$.
Thus, from the viewpoint of conventional electrodynamics, the ratio
\be
\Xi_{\vwen}
=
\frac{\vwenbar_\bot^{(e)}}{\vwenbar_\bot^{(m)}}
=
\frac{\av{\hat e}}{\av{\hat g}}
=
\frac{
1
+
\av{
\hat\sigma_3
}
}{
1
-
\av{
\hat\sigma_3
}
}
\ee
describes how much of the averaged wave energy would be stored in the electric component
as compared to the magnetic one if the medium has been replaced by a vacuum.

In case of a non-sustainable evolution (see Sec. \ref{s-nhgen-me}), these formulas stay intact except that the averages
must be computed with respect to the operator $\densnnorm$ not $\hat\rho$.\\

%\bc
\textit{(c) Transmitted power}.
%\ec
From the equations for the primary averages one can  
deduce an important quantum-statistical effect that occurs
during the EM wave's propagation in a medium.
In order to see this, let us introduce the total
transmitted power $\tpowerbar$
of the EM wave.
If the wave's evolution is sustainable, as defined in Sec. \ref{s-nhgen-me},
one can formally write that
\be
\tpowerbar
\equiv
\frac{1}{
%16 \pi
4}
%\overline{ 
\left\langle 
\int d x d y 
\left[
(\vee z\times\veeft^*)\cdot\vemft
+
\text{c.c.}
\right]
\right\rangle
,
\ee
where 
the bar denotes the average taken over
the statistical
ensemble represented by the normalized
density operator $\hat\rho$ (in case of sustainable evolution)
or the non-normalized
density operator $\densnnorm$ (in case of non-sustainable evolution). 
%the integral is taken over the medium's cross-section, as per usual.

In the latter case, 
we can
define 
a value
\be
\tpowerbar'
=
\frac{{\cal N}^2}{
%16 \pi
4}
\avo{\sigma_2}
,
\ee
by analogy to the single-mode power (\ref{e:onemodetpower}).
One can derive that
\be\lb{e:tompowerate}
\frac{d}{d \t}
\tpowerbar'
=
-
\frac{i}{
%16 \pi
4}
{\cal N}^2
\avo{\hamod - \hamod^\dagger}
,
\ee
so its right-hand side contains the expected
term which
vanishes in case of real permittivity and permeability,
cf. Eq. (\ref{e:pseudoherm}).
Therefore, $\tpowerbar'$ would be conserved
in case of real-valued $\varepsilon$ and $\mu$.

Furthermore,
in case of a sustainable evolution,
using 
%App. \ref{a:op} and 
Eq. (\ref{e:primav}), one can similarly define
\be\lb{e:tpower}
\tpowerbar
=
\frac{{\cal N}^2}{
%16 \pi
4}
\av{\sigma_2}
,
\ee
such that
$
\tpowerbar
=
\tpowerbar'
/
\text{tr} \densnnorm
$.
Using Eq. (\ref{e:primaveq2}), one obtains
the rate of power distribution along the propagation direction:
\be\lb{e:tpowerate}
\frac{d}{d \t}
\tpowerbar
=
2
\av{\hat\Gamma}
\tpowerbar
-
\frac{i}{
%16 \pi
4}
{\cal N}^2
\av{\hamod - \hamod^\dagger}
,
\ee
where
\be
\av{\hat\Gamma}
=
\av{\decoO}
+
\gamman
=
\frac{i}{2}
\av{
\hat\sigma_2 \hamod
-
\hamod^\dagger \hat\sigma_2
}
+
\gamman
\ee
is an average value of the decay operator (\ref{e:hamtoAH}).
In the $L$-representation,
Eq. (\ref{e:tpowerate}) reads
\ba
\frac{d}{d \t}
\tpowerbar
&=&
%\nn\\&&
2
\left(
\gamman
- \evre_- \av{\sigma_1}
+ 
%\evim_- \av{\sigma_2}
\frac{
%16 \pi
4}{{\cal N}^2}
\evim_-
\tpowerbar
\right)
\tpowerbar
\nn\\&&
-
\frac{1}{
%8 \pi
2}
{\cal N}^2
\left(
\evim_-
\av{\sigma_3}
+
\evim_+
\right)
,
\lb{e:tpowerateL}
\ea
for which derivation we have used Eqs. (\ref{e:hamoddefII}), (\ref{e:avGammapL}) and (\ref{e:tpower}).
One can see that
the right-hand side of Eq. (\ref{e:tpowerate}) contains the expected term
(proportional to $\av{\hamod - \hamod^\dagger}$) 
but it also contains the additional term
$2
\av{\hat\Gamma}
\tpowerbar$
which can remain non-zero even if both $\varepsilon$ and $\mu$
are real.
This term describes 
the purely quantum-statistical nonlinear effect --
the additional channel of energy's gain or loss (depending on a sign of $\hat\Gamma$)
that occurs during the sustainable wave evolution.
This effect
is yet another manifestation of
the sustainability-supporting energy flow,
described by
the last term
in Eq. (\ref{e:eomrho}).
In case of a weakly coupled environment, the magnitude of this effect must
be small but nevertheless viable for quantum EM devices and precise measurements.

To summarize, the transmitted power's behavior is different for two types of evolution 
discussed in Sec. \ref{s-nhgen-me} above.
%(Eq. (\ref{e:onemodetpower}) vs. (\ref{e:onemodetpower}))
This opens the possibility to determine experimentally
which type of evolution happens in a given physical configuration.

\sscn{Correlation functions}{s-nhgen-corr}

Here we consider only the case when wave evolution is described by the normalized density
operator, the other type can be easily considered by analogy.

Apart from the plain averages of operators (\ref{e:defmean}) taken in one point of $z$,
it is often necessary to consider correlations between different
or the same operators evaluated in the two or more points $\t_n > ... > \t_1 \geqslant \t_0$,
which was done in general NH case in Ref. \cite{sz14}.
In case of a two-point correlation function,
the definition adapted for our purposes would be
%\bw
\ba
{\cal C}_{\xi \chi} (\t_1, \t_2)
&\equiv&
\left\langle 
\chi (\t_2) \xi (\t_1)
\right\rangle
%\nn\\&=& 
= 
{\rm tr}\left\{\hat{\chi}
{\cal K}(\t_2, \t_1) \hat{\xi} \hat\rho (\t_1) \right\}
\nn\\&=& 
 {\rm tr}\left\{\hat{\chi}
{\cal K}(\t_2, \t_1)
\hat{\xi} {\cal K}(\t_1, \t_0)\hat\rho (\t_0)
\right\}
,
\label{e:herm-corf-nh}
\ea
%\ew
where $\hat{\chi}$ and $\hat{\xi}$ are operators in the Schr\"odinger representation,
and ${\cal K}$ is the (generalized) evolution operator defined as follows.
When ${\cal K}(\t_b, \t_a)$ is applied to anything on its right,
it evolves it from the point $\t_a$ to the point $\t_b$ using Eq. (\ref{e:eomrho}).
Hence, in the expression above, the first application
of ${\cal K}$ evolves $\hat\rho$ from the point $\t_0$ to $\t_1$
as a solution of (\ref{e:eomrho}). 
The second application of the evolution operator 
acts on the operator $\hat{\xi} \hat\rho(\t_1)$
and propagates it from the initial point $\t_1$ until the final point
$\t_2$, using Eq. (\ref{e:eomrho}).
Equation (\ref{e:herm-corf-nh})
has two obvious properties: it reduces to the 
conventional definition of a
correlation function
%of Hermitian quantum mechanics 
when $\hat{\Gamma}=0$, and to
the normalized average of $\hat{\chi}$, given by Eq. (\ref{e:defmean}), when $\hat{\xi}$ is
the identity operator.

Thus, the definition (\ref{e:herm-corf-nh}) 
is based on the spatial distribution of
the density matrix 
governed by Eqs. (\ref{e:dotTrOmega}) and (\ref{e:rhonorm}), or (\ref{e:eomrho}).
Naturally, the non-linearity of the latter
% equation (\ref{e:eomrho})
may invalidate the properties of the correlation function, which are
related to linearity.
Moreover,   
the linearizing ansatz (\ref{e:rhonorm}), often adopted in calculations,
can be applied only if the input of the evolution operator $\cal K$
has a unit trace.
Otherwise, one should use other analytical (or numerical) approaches.

The generalization of the definition of correlation functions 
(\ref{e:herm-corf-nh}) 
to a multi-point case is straightforward.
First, we introduce the ordered sets of third-axis coordinates 
$\t_n > ... > \t_1 \geqslant \t_0$ and
$s_m > ... > s_1 \geqslant \t_0$,
as well as their ordered union $\left\{\tau\right\}$:
$\tau_u > ... > \tau_1 \geqslant \t_0$ where $u \leqslant n+m  $.
Next, for any set of operators 
$\hat\chi_j$ ($j = 1,..., m$) and $\hat\xi_k$ ($k = 1,..., n$) in the
Schr\"odinger picture, one can define the superoperator 
$\Pi_l$ ($l=1,..., u$), cf. Refs. \cite{bpbook,gzbook}, through its action upon the operator $\hat D$ :
%\bw
\be
\Pi_l \hat D
= 
\left\{
\baa{lll}
\hat\xi_k \hat D & \text{if} \ \ \tau_l = \t_k \not= s_j  &  \text{for some} \ k \ \text{and all} \ j ,\\
\hat D \hat\chi_j & \text{if} \ \ \tau_l = s_j \not= \t_k &  \text{for some} \ j \ \text{and all} \ k ,\\
\hat\xi_k \hat D \hat\chi_j & \text{if} \ \ \tau_l = \t_k = s_j  &  \text{for some} \ k \ \text{and} \ j ,
\eaa
\right.
\ee
and for our case the operator $\hat D$ would be equal to $\hat\rho$.
Then 
%in the case of the definition (\ref{e:herm-corf-nh}),
one can define the multi-point correlation functions in the standard way:
\bw
\ba
{\cal C} (\t_1,..., \t_n; s_1,... , s_m) 
&\equiv&
%\equiv
\left\langle 
\chi_1 (s_1) ... \chi_m (s_m) \xi_n (\t_n) ... \xi_1 (\t_1)
\right\rangle
\nn\\&=&
\text{tr}
\left\{
\Pi_u {\cal K} (\tau_u, \tau_{u-1})
\Pi_{u-1} {\cal K} (\tau_{u-1}, \tau_{u-2})
...
\Pi_{1} {\cal K} (\tau_{1}, \t_{0})
\hat\rho (\t_0)
\right\}
,
\label{e:mcorf-nh}
\ea
\ew
where the evolution operator $\cal K$ is defined above.

\sscn{Entropy}{s-nhgen-entr}

Let us consider first the sustainable case -- when wave evolution is described by the normalized density
operator.
Apart from purity $\text{tr} \hat\rho^2$ and linear entropy $S_L = 1 -\text{tr}\hat\rho^2$,
there exists another characteristic value describing the amount of disorder
and statistical uncertainty in a system -- the quantum entropy of the Gibbs type.
In Ref. \cite{sz15} it was shown that for a system driven by
NH Hamiltonian one can introduce two types of quantum entropy:
the conventional Gibbs-von-Neumann one
\be
S_{\rm vN}
\equiv
-k_B \left\langle   \ln\hat{\rho} \right\rangle
=
-k_B{\rm tr}\left( \hat{\rho} \ln\hat{\rho} \right)
,
\label{e:SVNnH}
\ee
and the NH-adapted Gibbs-von-Neumann one
\begin{equation}
S_\text{NH}
\equiv
-k_B \langle   \ln\densnnorm \rangle
=
-k_B{\rm tr} ( \hat{\rho} \ln\densnnorm ) 
= 
- k_B 
\frac{
{\rm tr}( \densnnorm \ln\densnnorm )
     }{
{\rm tr}\, \densnnorm
}
,
\label{e:SVNnH2}
\end{equation}
where $k_B$ is the Boltzmann constant.
The two notions of entropy are related by the formula
\be\lb{e:diffentr}
S_\text{NH}
=
S_{\rm vN}
-
k_{\rm B}
\ln{( \text{tr}\, \densnnorm)}
,
\ee
therefore, the difference between $S_\text{NH}$ and $S_{\rm vN}$ 
is a measure of deviation of $\text{tr}\, \densnnorm$
from unity.  
One can see that the entropy $S_\text{NH}$ combines both
the normalized and ``primordial'' (non-normalized) density
operators, and thus can signal
the expected thermodynamic behavior
of an open system.
The entropy $S_\text{NH}$ also
seems to be more suitable for describing the gain-loss processes that
are related to the non-conservation of entire probability sample space measure, since
it contains information not only about
the conventional von Neumann entropy but
also about the trace of the operator $\densnnorm$,
according to the relation (\ref{e:diffentr}).
Assuming that $S_\text{vN}$ is bound at large times,
the NH entropy grows when $\text{tr}\,\densnnorm$
decreases, also it takes positive values if $\text{tr}\,\densnnorm < 1$
and negative ones otherwise.
Hence,
one can say that in our case $S_\text{NH}$ 
takes into account the statistical uncertainty, which comes from
the flow of EM energy between the wave and its environment.

As for the case of non-sustainable evolution, governed by the non-normalized density
operator, one can only define the Gibbs-von Neumann entropy
%of the form
\begin{equation}
S_\text{NH}^\prime
\equiv
-k_B \avo{\ln\densnnorm}
=
-k_B{\rm tr} ( \densnnorm \ln\densnnorm ) 
,
\end{equation}
and the linear entropy $S_L^\prime = 1 -\text{tr}\densnnormsq$.

\sscn{Hamiltonian ``gauge'' transformations}{s-nhgen-gau} 

One can see that the last term $\hamto_{\cal N}$ 
of the Hamiltonian operator (\ref{e:hamto})
is proportional to the identity operator, therefore,
one could wonder what kind of physics can be described by such terms.
%In order to find the answer and
Expanding the discussions presented in Refs. \cite{sz13,sz14,sz15}, let us
consider
the following ``shift'' transformation 
of the decay operator
\be\lb{e:transgfau}
\trans{\hat\Gamma}
\mapsto
\hat\Gamma = \trans{\hat\Gamma} + \frac{1}{2}\hbarr \alpha (z) \hat{\cal I}
,
\ee
where $\alpha (z)$ is an arbitrary real-valued function.
% and $\hat I$ is the unity operator.
This transformation is similar to the transformation
\be
\trans{\hat{\cal H}}
\mapsto
\hat{\cal H} = \trans{\hat{\cal H}} + c_0 \hat{\cal I}, 
\label{e:H-const-shift}
\ee
$c_0$ being an arbitrary complex number, which is the non-Hermitian generalization
of the energy shift in conventional quantum mechanics.
Therefore, in Refs. \cite{sz13,sz14} it was called
the ``gauge'' transformation of the Hamiltonian,
whereas the terms of the type $c_0 \hat{\cal I}$ can be called the ``gauge'' terms.

By direct substitution one can  show that the equation (\ref{e:eomrho}) is invariant under the
transformation (\ref{e:transgfau}),
therefore, one immediately obtains
\be
\hat\rho = \trans{\hat\rho}, \
S_\text{vN} = \trans{S_\text{vN}}
,
\ee
therefore the von Neumann entropy is not affected
by the transformation 
%in Eq. 
(\ref{e:transgfau}).
One can see that any information regarding the
``shift'' of the total non-Hermitian Hamiltonian 
is lost if one deals solely with the normalized density operator.

However, equation (\ref{e:dotTrOmega}) is not
invariant under the transformation (\ref{e:transgfau}).
If both $\hamto_+$ and $\hat{\Gamma}$ commute with $z$ 
%(as they do in our case)
then,
substituting (\ref{e:transgfau}) into (\ref{e:dotTrOmega}),
we obtain that the non-normalized density acquires an exponential 
factor:
\be
\densnnorm = 
\trans{\densnnorm} 
\exp{\left(-\int\limits_0^z \!\alpha (\zeta)\, d \zeta \right)},
\ee
%where by integral we mean the primitive.
such that,
recalling Eq. (\ref{e:diffentr}), one obtains
\be
S_\text{NH} 
= 
\trans{S_\text{vN}} - k_B \ln{\text{tr}\, \densnnorm}
=
\trans{S_\text{NH}} + k_B \int\limits_0^z \!\alpha (\zeta)\, d \zeta
,
\ee
which indicates that any information 
about the ``shift'' term in Eq. (\ref{e:H-const-shift})
in the total non-Hermitian Hamiltonian, which was
lost during the normalization
procedure (\ref{e:rhonorm}),
can be recovered by means of the NH entropy.

\scn{Two-level models}{s-tls}

As one can see from Sec. \ref{s-ana} and Appendix \ref{a:op},
the features of two-level systems (TLS) have sufficiently tight links with the EM
wave's propagation inside media.
Thus, in our case the TLS approach is
not just an approximation, but a simple way to construct the models
that reflect main symmetries and statistical properties of a full theory,
yet they are not too complex to be studied analytically.
In essence, this approach
is responsible for describing the physics of the energy
transfers between electric and magnetic components
of EM wave, which propagates in the medium and interacts with its environment.
% in presence of quantum-statistical effects.
Indeed, from the $L$-representation, described
in Sec. \ref{s-ana-ev},
one can see that
the features of the electrical and magnetic components
of each mode are separately encoded in the
eigenvalues $\eigv e$ and $\eigv m$ (which can differ from mode
to mode, and can be complex-valued
in general), therefore,
an appropriate two-level model would describe quantum transitions between them,
as well as any related (quantum-)statistical and
dissipative effects that may occur.

\sscn{Generic model with constant parameters}{s-tls-gen} 

With the use of Sec. \ref{s-ana-ev},
for a given mode, in the $L$-representation, we can write the following
Hamiltonian
%\bw
\ba
&&
\hamto
%= \hamo + \hamto_{\cal N}
=
%\hbarr \hat\sigma_2 \hamod
\hamo
-
i 
%\hbarr
\gamman
\hat I
=
i
\twomat{-\gamman}{-\eigv{m}}
{\eigv{e}}{-\gamman}
,
\lb{e:hamtoIItls}
\\
&&
\hamo
=
\hat\sigma_2 \hamod
=
i \eigv{-}
\hat\sigma_1
+
\eigv{+}
\hat\sigma_2
=
\twomat{0}{-i\eigv{m}}
{i\eigv{e}}{0}
,~~~
\ea
where $\hat\sigma$'s are the standard $2\times 2$ Pauli matrices,
and
\be\lb{e:hamtodIItls}
\hamod
=
\eigv{-}
\hat\sigma_3
+
\eigv{+}
\hat{I}
=
\twomat{\eigv{e}}{0}
{0}{\eigv{m}}
,
\ee
and  the $\lambda$'s notations of Sec. \ref{s-ana-ev} are implied.
The eigenvalues $\eigv{e,m}$ become $\t$-independent
if one assumes 
%from now on
that the permittivity and permeability are functions
of the transverse coordinates only.
It means that waveguide's material possesses, at least, one translational symmetry --
along 
the direction of wave's propagation (\textit{i.e.}, $\t$-axis),
which is a case for a large class of
optical fibers, long scatterers of constant cross-section, and integrated, 
nanophotonic and plasmonic waveguides.
Note that in general, coefficient $\gamman$, as well as ${\cal N}$, can be a function
of $\t$, according
to Eqs. (\ref{e:wfnorm}) and (\ref{e:gamman}).
However, because EM wave's propagation inside the above-mentioned materials results in the fields' dependence on $\t$
being of an exponential type, the integral (\ref{e:wfnorm}) becomes dominated by an exponential function
of $\t$; therefore the coefficient $\gamman$ can be assumed constant in the leading approximation (in many cases,  exactly constant or even zero).
Thus, here we assume the components of the matrix (\ref{e:hamtoIItls}) being constant but otherwise free parameters (their specific values can be always found for a given mode).

As in Sec. \ref{s-nhgen-me},
the Hamiltonian (\ref{e:hamtoIItls})
can be easily decomposed into self-adjoint and skew-adjoint parts
to acquire the form (\ref{e:nhham}),
with
\bse\lb{e:hamtoHaHtls}
\ba
\hamto_+
&=&
{\hamo}_+
=
\frac{1}{2}
\left(
\hat\sigma_2 \hamod + \hamod^\dagger \hat\sigma_2
\right)
%\nn\\&=&
=
\evim_- \hat\sigma_1
+
\evre_+ \hat\sigma_2
\nn\\&
=&
\twomat{0}{- \frac{i}{2} (\eigv{e}^* +\eigv m)}{\frac{i}{2}(\eigv{e} + \eigv{m}^*)}{0}
,\lb{e:hamtoHtls}\\
\hat\Gamma
&=&
\decoO
+
\gamman
\hat{\cal{I}}
=
\frac{i}{2}
\left(
\hat\sigma_2 \hamod - \hamod^\dagger \hat\sigma_2
\right)
+
%\pDer z \!\left(\ln{{\cal N}}\right)
\gamman
\hat{{I}}
\nn\\&
=&
- \evre_- \hat\sigma_1
+
\evim_+ \hat\sigma_2
+
\gamman
\hat{{I}}
\nn\\&
=&
\twomat{\gamman}{- \frac{1}{2} (\eigv{e}^* -\eigv m)}
{-\frac{1}{2}(\eigv{e} - \eigv{m}^*)}{\gamman}
,
\lb{e:hamtoAHtls}
\ea
\ese
where the notations 
(\ref{e:coeffRI})
% and (\ref{e:coeffI})
are assumed.\\

%\bc
\textit{(a) Eigenvalues and singular points}.
%\ec
The eigenvalues of the Hamiltonian (\ref{e:hamtoIItls})
are
\be\lb{e:evhamtoIItls}
\Lambda_\pm
=
\pm\sqrt{\eigv e} \sqrt{\eigv m}
- i \gamman
,
\ee
therefore, the ``energies'' and ``resonance'' (half-)widths 
(we use quotes because of working within the framework of the Maxwell-Schr\"odinger analogy)
are given, respectively,
by
\bse
\ba
&&
E_\pm \equiv \text{Re} (\Lambda_\pm) = 
\pm \text{Re}\left( \sqrt{\eigv e} \sqrt{\eigv m} \right)
,\\&&
\Gamma_\pm \equiv \text{Im} (\Lambda_\pm) = 
- \gamman \pm \text{Im} \left(\sqrt{\eigv e} \sqrt{\eigv m} \right)
,~~
\ea
\ese
such that $\Lambda_\pm = E_\pm + i \Gamma_\pm$.

The necessary condition for a singular point (SP) is when
the two eigenvalues coalesce (if corresponding eigenfunctions also coincide then
such a point called exceptional).
For our model this results in the following
condition:
\be
\eigv{e} \eigv m = 0
,
\ee
which is equivalent to
two relations
\bse
\ba
&&
\text{Re} (\eigv{e}) \text{Re} (\eigv{m})
-
\text{Im} (\eigv{e}) \text{Im} (\eigv{m})
=0
,\\&&
\text{Im} (\eigv{e}) \text{Re} (\eigv{m})
+
\text{Re} (\eigv{e}) \text{Im} (\eigv{m})
=0
,
\ea
\ese
where the formulas (\ref{e:coeffReeImm})
%-(\ref{e:coeffImm}) 
are implied.
These relations narrow the set of all SP-compatible eigenvalues' components,
\be
\{ \lambda \}_{c}=
\left\{\text{Re} (\eigv{e}),\, \text{Im} (\eigv{e}),\,\text{Re} (\eigv{m}),\,\text{Im} (\eigv{m}) \right\}
,
\ee
down to seven possible combinations
\bse
\ba
&&
\{ \lambda \}_{c}^{(0)}=
\left\{0,\,0,\,0,\,0 \right\}
,
\\&&
\{ \lambda \}_{c}^{(\text{Ia})}=
\left\{\text{Re} (\eigv{e}),\, 0,\,0,\,0 \right\}
,
\\&&
\{ \lambda \}_{c}^{(\text{Ib})}=
\left\{0,\,\text{Im} (\eigv{e}),\, 0,\,0 \right\}
,
\\&&
\{ \lambda \}_{c}^{(\text{Ic})}=
\left\{0,\,0,\,\text{Re} (\eigv{m}),\, 0 \right\}
,
\\&&
\{ \lambda \}_{c}^{(\text{Id})}=
\left\{0,\,0,\,0,\, \text{Im} (\eigv{m}) \right\}
,
\\&&
\{ \lambda \}_{c}^{(\text{IIa})}=
\left\{\text{Re} (\eigv{e}),\, \text{Im} (\eigv{e}),\,0,\,0 \right\}
,
\\&&
\{ \lambda \}_{c}^{(\text{IIb})}=
\left\{0,\,0,\,\text{Re} (\eigv{m}),\, \text{Im} (\eigv{m}) \right\}
,
\ea
\ese
where the non-zero components can be any real numbers.
Thus, this results in the simplest classification of eigenvalues $\eigv{e,m}$
which tells us that for a singular point to exist, at least one of the eigenvalues must vanish.
%(for instance, vanishing of $\lambda_e$ happens for a longitudinal bulk plasmon \cite{ref16}).
In other words, our system (\ref{e:hamtoIItls})
avoids level crossing unless at least one of its matrix off-diagonal components vanishes.\\

%\bc
\textit{(b) Density matrix}.
%\ec
Due to the completeness of Pauli spin operators in two-dimensional Hilbert space,
the exact solutions of the master equations for the
density operators of the system
can be searched in the form
\ba
&&
\densnnorm
=
\frac{1}{2}
\left(
\text{tr} \densnnorm\, \hat I
+
\sum\limits_{k=1}^3
\avo{\sigma_k}\, \hat\sigma_k
\right)
,\lb{e:decompom}\\&&
\hat\rho
\equiv
\frac{\densnnorm}{\text{tr} \densnnorm }
=
\frac{1}{2}
\left(
\hat I
+
\sum\limits_{k=1}^3
\av{\sigma_k} \hat\sigma_k
\right)
,
\lb{e:decomprho}
\ea
where the averages $\avo{\sigma_k}$ and $\av{\sigma_k} $ satisfy Eqs.
(\ref{e:primavoLeqL}) and (\ref{e:primaveqL}), respectively.
Notice that Eqs.  (\ref{e:primaveqL}), hence
the normalized density $\hat\rho$ and related mean values, do not depend on $\gamman$,
as discussed in Sec. \ref{s-nhgen-gau}.

Further, Eqs. (\ref{e:primavoLeqL}) are primary equations to solve,
and they must be supplemented with the ``initial'' (boundary) conditions.
Following Sec. \ref{s-nhgen-ini}, we impose:
\be\lb{e:doini}
%\hat\rho (\t =0) 
%|_{\t = 0} = 
\densnnorm 
%(\t =0) 
|_{\t = 0}
= \densnnorm_0
,
\ee
where the matrix
$\densnnorm_0$ describes the state which corresponds to the 
EM wave at the boundary of a medium (assuming that the latter occupies
the region $\t \geqslant 0$).

For instance, one can assume that the wave is originally (an instant before entering a medium)
%polarized (circularly or linearly), i.e.,
in a pure state with respect to the density $\densnnorm$.
This means that the matrix
$\densnnorm_0$ must be idempotent:
\be\lb{e:idemptls}
\densnnormOsq = \densnnorm_0 
.
\ee
However, as mentioned in Sec. \ref{s-nhgen-ini}, there is still a certain ambiguity
because the trace of $\densnnorm_0$ does not necessarily equal to one.
Therefore, one must differentiate the following two cases:

(1) $\text{tr} \densnnorm_0 = 1$.
Then this  operator can
be parametrized using the Bloch sphere (such a parametrization is also of interest
in the coupled mode theory \cite{che03}):
\ba
\densnnormOo
&=&
%\left|\Psi_0 \right\rangle\!\left\langle \Psi_0 \right| =
%\bcs
%\frac{1}{2} \twomat{1+\tilde\varrho}{\varrho_1 + i \varrho_2}{\varrho_1 - i \varrho_2}{1-\tilde\varrho} 
%\twomat{\frac{1}{2}(1-\bar\varrho)}{\varrho_1 + i \varrho_2}{\varrho_1 - i \varrho_2}{\frac{1}{2}(1+\bar\varrho)} 
%{0<\varrho_1<1 \choose{\varrho_2^2 < \varrho_1 - \varrho_1^2}} 0<\varrho_1<1
\twomat
{\Sin{2}{\theta_0/2}}
{\frac{1}{2}\text{e}^{i \phi_0}\sin\theta_0}
{\frac{1}{2}\text{e}^{-i \phi_0}\sin\theta_0}
{\Cos{2}{\theta_0/2}}
\nn\\&
=&
\frac{1}{2}
\left[
\hat I
+
\sin{\theta_0}
(
\cos{\phi_0} \,
\hat\sigma_1
-
%\sin{\theta_0}
\sin{\phi_0} \,
\hat\sigma_2
)
%\nn\\&&
-
%\frac{1}{2}
\cos{\theta_0} 
\hat\sigma_3
%\frac{1}{2}
\right]
,
\nn\\
\lb{e:rhobloch}
\ea
where we use the notations from Appendix \ref{a:bs}.
Notice that in this case 
\be
\hat\rho^{(1)}|_{\t = 0} 
= 
\densnnormOo / \text{tr}\, \densnnormOo
=
\densnnormOo
\ee
as well.

(2) $\text{tr} \densnnorm_0 \not= 1$.
One can easily show that the only non-trivial 2$\times$2 Hermitian matrix with a non-unit trace,
which satisfies the property (\ref{e:idemptls}), 
is the identity matrix:
\be\lb{e:rhoident}
\densnnormO
= \hat I
.
\ee
Notice that in this case 
the normalized operator
\be\lb{e:nrhoident}
\hat\rho^{(2)}|_{\t = 0} 
= 
\densnnormO
 / \text{tr} {\densnnormO}
=
\frac{1}{2}
\hat I
\ee
is neither equal to $\densnnormO$ nor idempotent (pure).
The latter means that $\hat\rho^{(2)}_0$ describes the mixed state -- in which
there is an equal probability to find the system in either of the basis
states $\hat e$ and $\hat g$ defined in Appendix \ref{a:bs}.

To summarize, for practical computations in the two-level models, one can choose
boundary conditions of either (\ref{e:rhobloch}) type or given by Eq. (\ref{e:rhoident}),
which correspond to the states, which are, respectively, either pure or classical-type mixtures
with respect to $\hat\rho$.

\sscn{Homogeneous medium with constant cross-section area and real frequency-independent permittivity and permeability}{s-tls-rea} 

Let us  now consider the special case of the model described 
in Sec. \ref{s-tls-gen}, for which 
 $\varepsilon$ and $\mu$ are real-valued constants:
\be
\mu = \text{const},
\
\varepsilon = n^2/\mu
= \text{const}
,
\ee
where $n = \sqrt{\varepsilon \mu}$ is a relative refractive index between the medium and physical vacuum.\\
%The total index of refraction would be thus equal to $n / \sqrt{\varepsilon_v \mu_v}$, $\varepsilon_v$ and $\mu_v$ being, respectively, the permittivity and permeability of the physical vacuum.\\

%\bc
\textit{(a) Hamiltonian}.
%\ec
In this case, the operators (\ref{e:oplem}) and (\ref{e:hamtodIItls})
are Hermitian, therefore constants $\eigv e$ and $\eigv m$
are real-valued.
Moreover, Eqs. (\ref{e:evoplem}) can be exactly solved
by means of the plane-wave ansatz.
When using the latter for a medium with a finite-size cross-section we assume that either
the cross-section area is large enough to neglect boundary near-field effects,
or the fields inside
the medium can be matched with the fields
outside it, for example those which decay at spatial infinity,
and one can find a solution of Eqs. (\ref{e:evoplem}) by means of the Fourier or Laplace transforms. 
This matching can be achieved by imposing suitable conditions across the medium's surface or interface, 
for instance one can assume the smoothness of fields across this surface.
Because we are dealing with the Maxwell-Schr\"odinger analogy, this matching would be somewhat similar
to the quantum-mechanical problem of a particle in a finite potential well.

One can derive that
\bse\lb{e:lamtlsRe}
\ba
&&
\eigv e
=
\varepsilon \omega - \frac{|\wavnt|^2}{\mu \omega}
=
\varepsilon \omega
\left[
1 - 
\left(
\frac{|\wavnt|}{n \omega}
\right)^2
\right]
,\\&&
\eigv m
=
\mu \omega - \frac{|\wavnt|^2}{\varepsilon \omega}
=
\mu \omega
\left[
1 - 
\left(
\frac{|\wavnt|}{n \omega}
\right)^2
\right]
,
\ea
\ese
where $|\wavnt| = \sqrt{k_x^2 + k_y^2} $ is the transverse wavenumber.
Correspondingly, the absolute value of the transverse phase velocity of wave
is
\be
v_p = \omega / |\wavnt|
.
\ee
It is convenient to express this wavenumber in terms of the frequency at which
the eigenvalues vanish:
\be
\omega_0 = |\wavnt|/n
,\
\eigv{e, m}|_{\omega = \omega_0} = 0
,
\ee
and rewrite Eqs. (\ref{e:lamtlsRe})
in the form:
\bse\lb{e:lamtlsRe2}
\ba
\eigv e
&=&
\varepsilon \om
,\\
\eigv m
&=&
\mu \om
,
\ea
\ese
and
\be
v_p = \tilde\omega /n
,
\ee
where 
we have denoted
\be
\om 
%= \om (\omega)
=
\omega
\left(
1 - 
%\left(
\frac{\omega_0^2}{\omega^2}
%\right)^2
\right)
=
\omega_0
\left(
\tilde\omega - 
%\left(
\frac{1}{\tilde\omega}
%\right)^2
\right)
,
\ee
and $\tilde\omega = \omega / \omega_0$.
%Notice that these relations are similar to those arising for the Drude plasmons if $\omega_0$ equals to the plasma frequency \cite{ref16}.
Quantity $\omega_0$ can be interpreted as the frequency at which, for a given $\wavnt$, there is no ``evolution'' along $z$-axis; this corresponds to a wave propagating normal to the $z$-axis in the $xy$ plane \cite{ref16}. 
In our plane-wave ansatz, this can be related to 
%what is called in EM 
the light-line limit, which is a tenet of the coupling between small-scale structures and the far-field, notably in all kinds of nano-objects and in bounded periodic media 
%For too small omega, such wavevector corresponds to an attenuated near-field component decaying or growing exponentially 
\cite{nhbook}.

Further, one can show that for this model the functional (\ref{e:wfnorm})
becomes:
\be
{\cal N}^2 
\to
{\cal N}^2_0 
=
%\left.
{\cal A}
\left(
\left|\veeft (0)\right|^2  + \left| \vemft (0)\right|^2
\right)
%\right|_{x=y=z=0}
,
\ee
where 
$\veeft (0) = \veeft|_{x=y=z=0}$, similarly for $\vemft$,
and
${\cal A}$ is the cross-section area.
If we assume the latter being independent of $\t$ then,
according to Eq. (\ref{e:gamman}),
\be
\gamman = 0
,
\ee
and 
the Hamiltonian has the form (\ref{e:nhham}),
with
\bse\lb{e:hamtoAHtlsRe}
\ba
&&
\hamto_+
=
%{\hamto'}_+ =
\frac{1}{2}
\left\{
\hat\sigma_2,\, \hamod 
%+ \hamod^\dagger \hat\sigma_2
\right\}
%\nn\\&=&
=
%\evim_- \hat\sigma_1 +
\eigv + \hat\sigma_2
%\nn\\&=&
=
%\twomat{0}{- \frac{i}{2} (\eigv{e} +\eigv m)}{\frac{i}{2}(\eigv{e} + \eigv{m})}{0}
\twomat{0}{- i \eigv{+}}{i \eigv{+}}{0}\!
,~~~
%\lb{e:hamtoHtlsR}\\&&
\\&&
\hat\Gamma
=
%\decoO
%+ \gamman \hat{\cal{I}}=
\frac{i}{2}
\left[
\hat\sigma_2,\, \hamod 
%- \hamod^\dagger \hat\sigma_2
\right]
%\pDer z \!\left(\ln{{\cal N}}\right)
%+ \gamman \hat{{I}}
%\nn\\&=&
=
- \eigv - \hat\sigma_1 
%+ \evim_+ \hat\sigma_2
%+ \gamman \hat{{I}}
%\nn\\&=&
=
%\twomat{\gamman}{- \frac{1}{2} (\eigv{e} -\eigv m)}{-\frac{1}{2}(\eigv{e} - \eigv{m})}{\gamman}
%\twomat{\gamman}{- \eigv{-}}{-\eigv{-}}{\gamman}
\twomat{0}{- \eigv{-}}{-\eigv{-}}{0}\!,
\ea
\ese
where Eqs. (\ref{e:coeffRI})
% and (\ref{e:coeffI})
have been used,
and 
\ba
\eigv{\pm} &=& 
\frac{1}{2} 
h_\pm
\om
%\nn\\&=&
,
\lb{e:lampmhomo}
\ea
where $h_\pm = \varepsilon \pm \mu$.

\begin{figure}[t]
\centering
\subfloat[$\omega_0 \not= 0$]{
  \includegraphics[width=0.49\columnwidth]{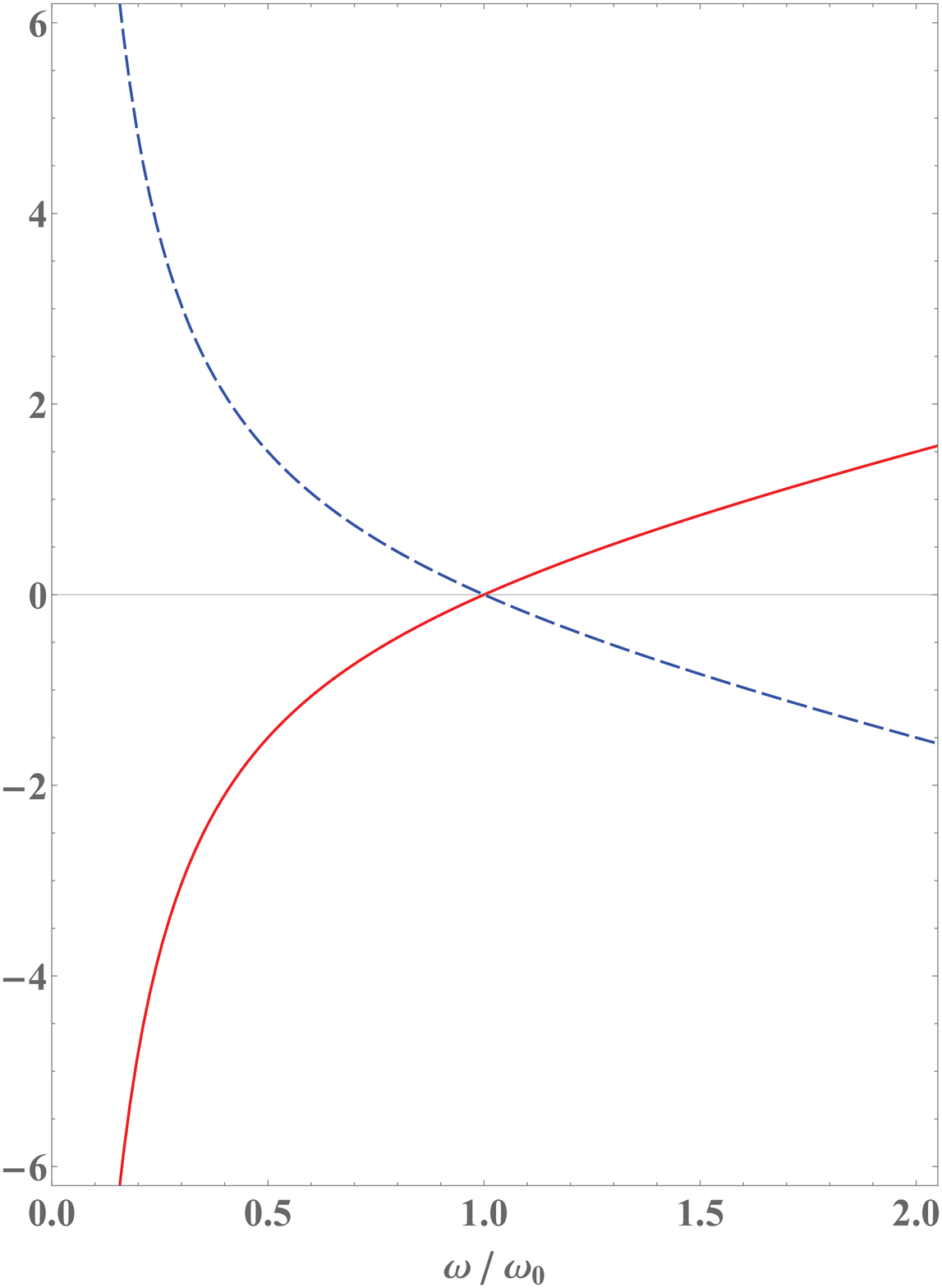}
}
%\hspace{3mm}
\subfloat[$\omega_0 = 0$]{
  \includegraphics[width=0.49\columnwidth]{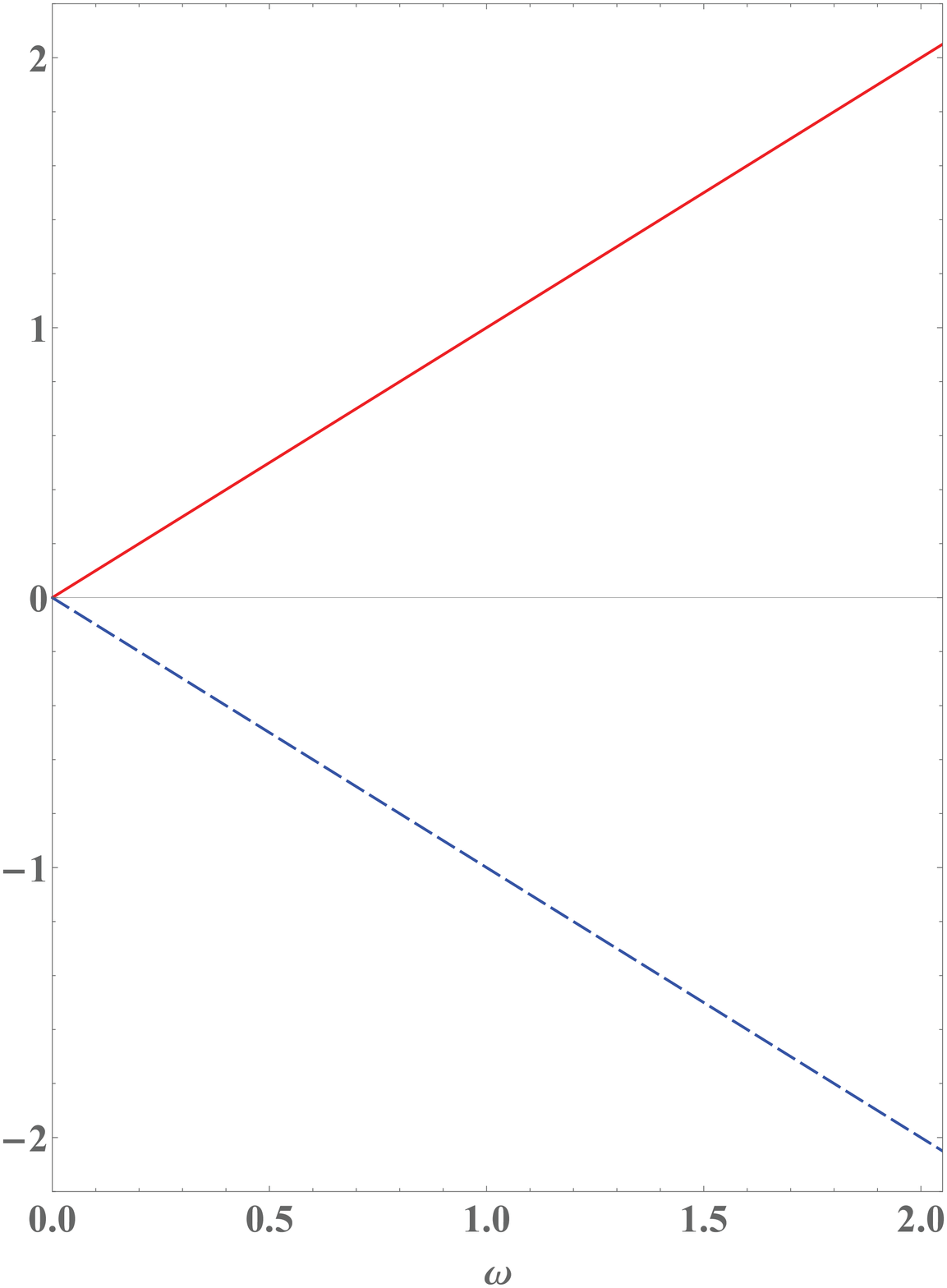}
}
\hspace{0mm}
\caption{Eigenvalues (\ref{e:evhamtoIItlsRe}) 
versus frequency.
The vertical axis's units are $n \omega_0$ (left panel)
or $n \times \text{Hz}$ (right panel).
Solid curves correspond to $\Lambda_+$, dashed ones -- to $\Lambda_-$.
%The initial state is given by.
}
\label{f:tlsrehev}
\end{figure}

Further, according to Eq.  (\ref{e:evhamtoIItls}),
the eigenvalues of the Hamiltonian (\ref{e:hamtoAHtlsRe})
are
\be\lb{e:evhamtoIItlsRe}
\Lambda_\pm
=
%E_\pm =
%\pm\sqrt{\eigv e \eigv m} =
\pm 
n \om
=
\pm
n
\omega
\left(
1 - 
%\left(
\omega_0^2 / \omega^2
%\right)^2
\right)
,
\ee
from which one can see that the 
levels cross at $\omega = \omega_0$, where both eigenvalues vanish.
Thus, for $\omega_0 \not= 0$,
%and permittivity and permeability being $\omega$-independent, 
the Hamiltonian (\ref{e:hamtoAHtlsRe})
becomes singular at $\omega =0$ 
and vanishes at  $\omega = \omega_0$.
At $\omega_0 = 0$,
% and permittivity and permeability being $\omega$-independent, 
the energy becomes proportional to the
frequency, similarly to the quantum harmonic oscillator, see Fig. \ref{f:tlsrehev}.\\
%Note that this picture can drastically change if permittivity and permeability themselves become depending on $\omega$.\\

\begin{figure}[t]
\centering
\subfloat[$\omega_0 \not= 0$]{
  \includegraphics[width=0.49\columnwidth]{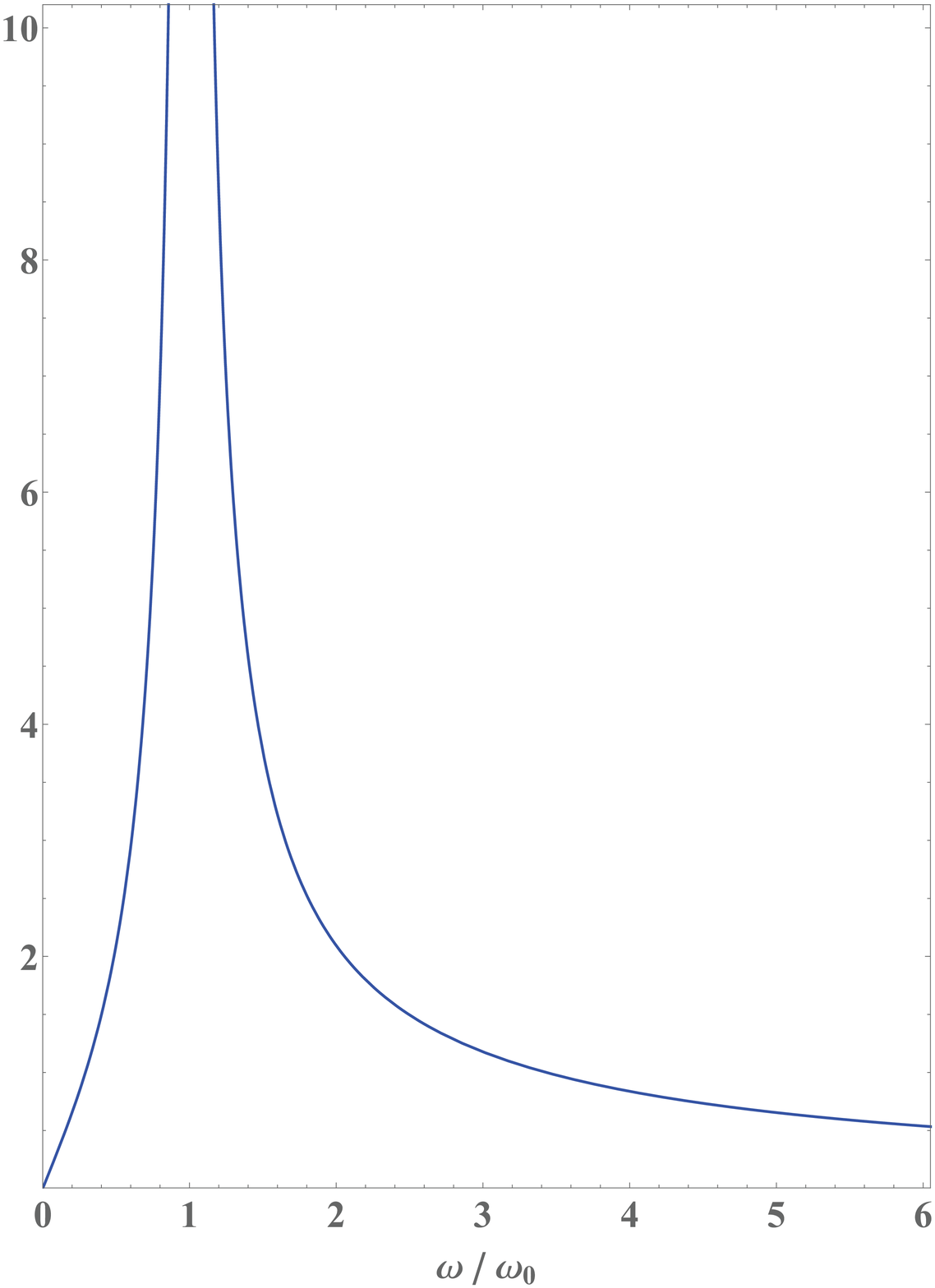}
}
%\hspace{3mm}
\subfloat[$\omega_0 = 0$]{
  \includegraphics[width=0.49\columnwidth]{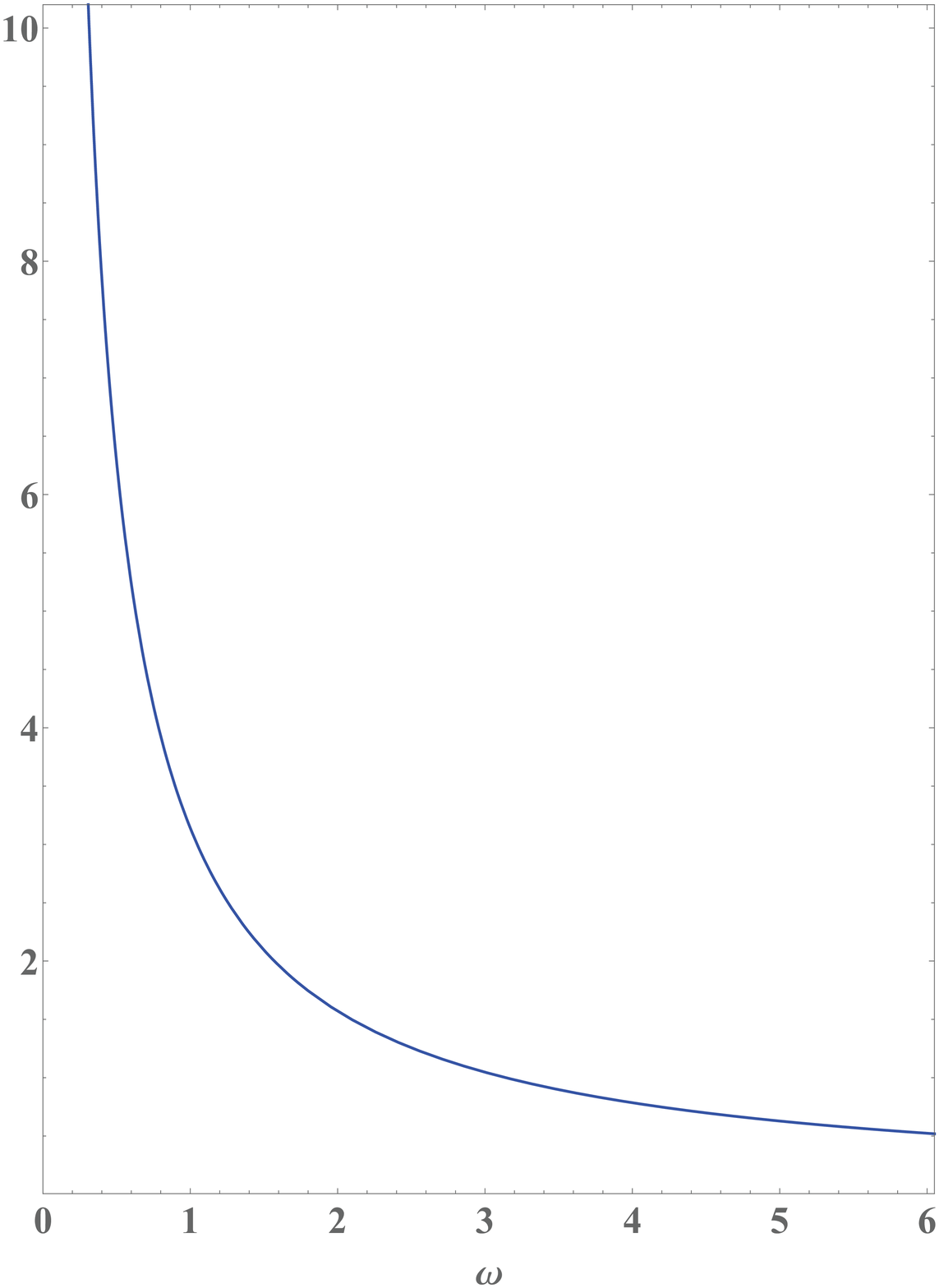}
}
\hspace{0mm}
\caption{Period of oscillations (\ref{e:tlsreperiod}) 
versus frequency.
The vertical axis's units are $(n \omega_0)^{-1}$ (left panel)
or $(n \times\text{Hz})^{-1}$ (right panel).
%Solid curves correspond to $\Lambda_+$, dashed ones -- to $\Lambda_-$.
%The initial state is given by.
}
\label{f:tlsreperiod}
\end{figure}

\begin{figure}[t]
\centering
\subfloat[$\omega_0 \not= 0$]{
  \includegraphics[width=0.49\columnwidth]{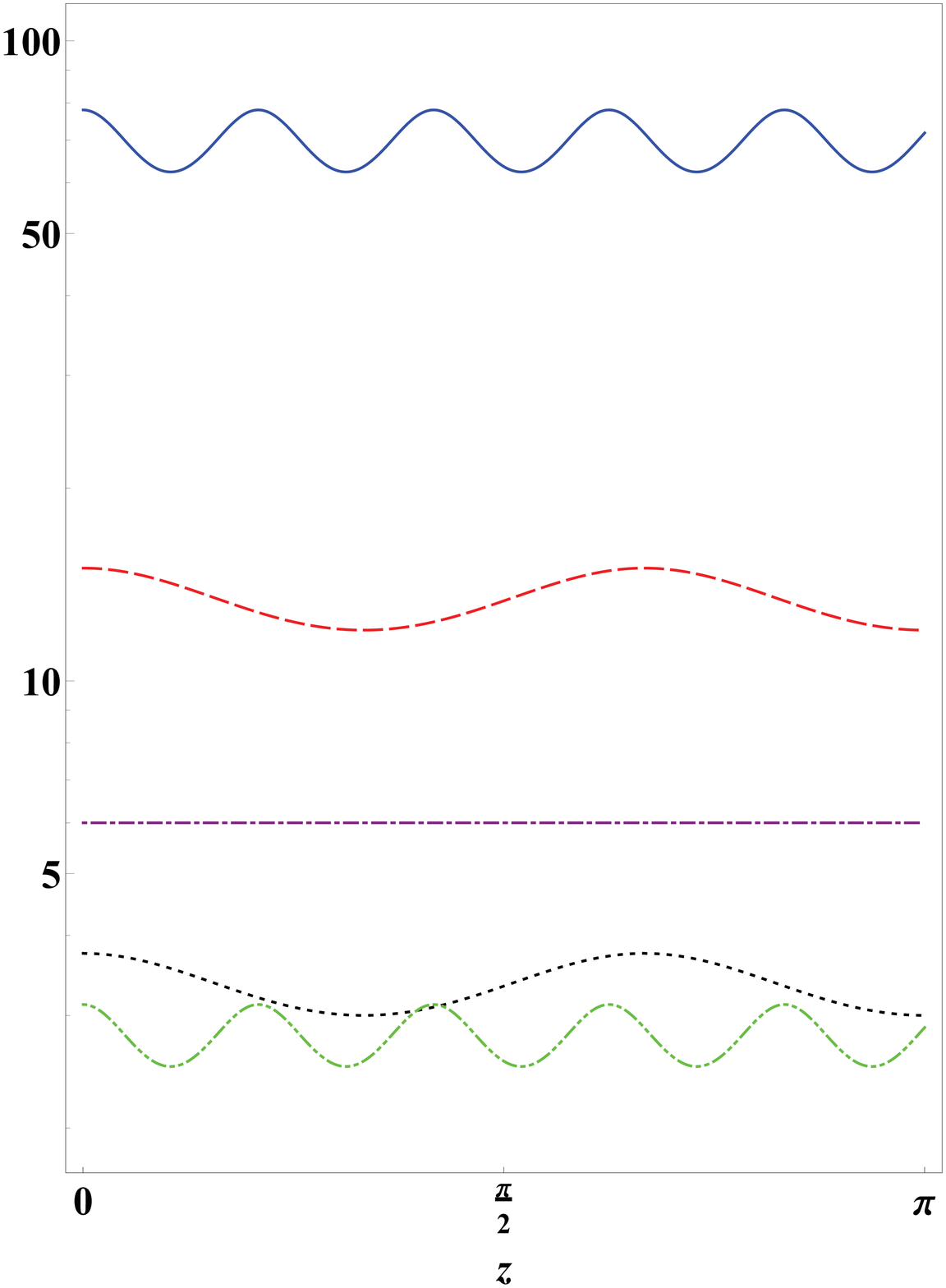}
}
%\hspace{3mm}
\subfloat[$\omega_0 = 0$]{
  \includegraphics[width=0.49\columnwidth]{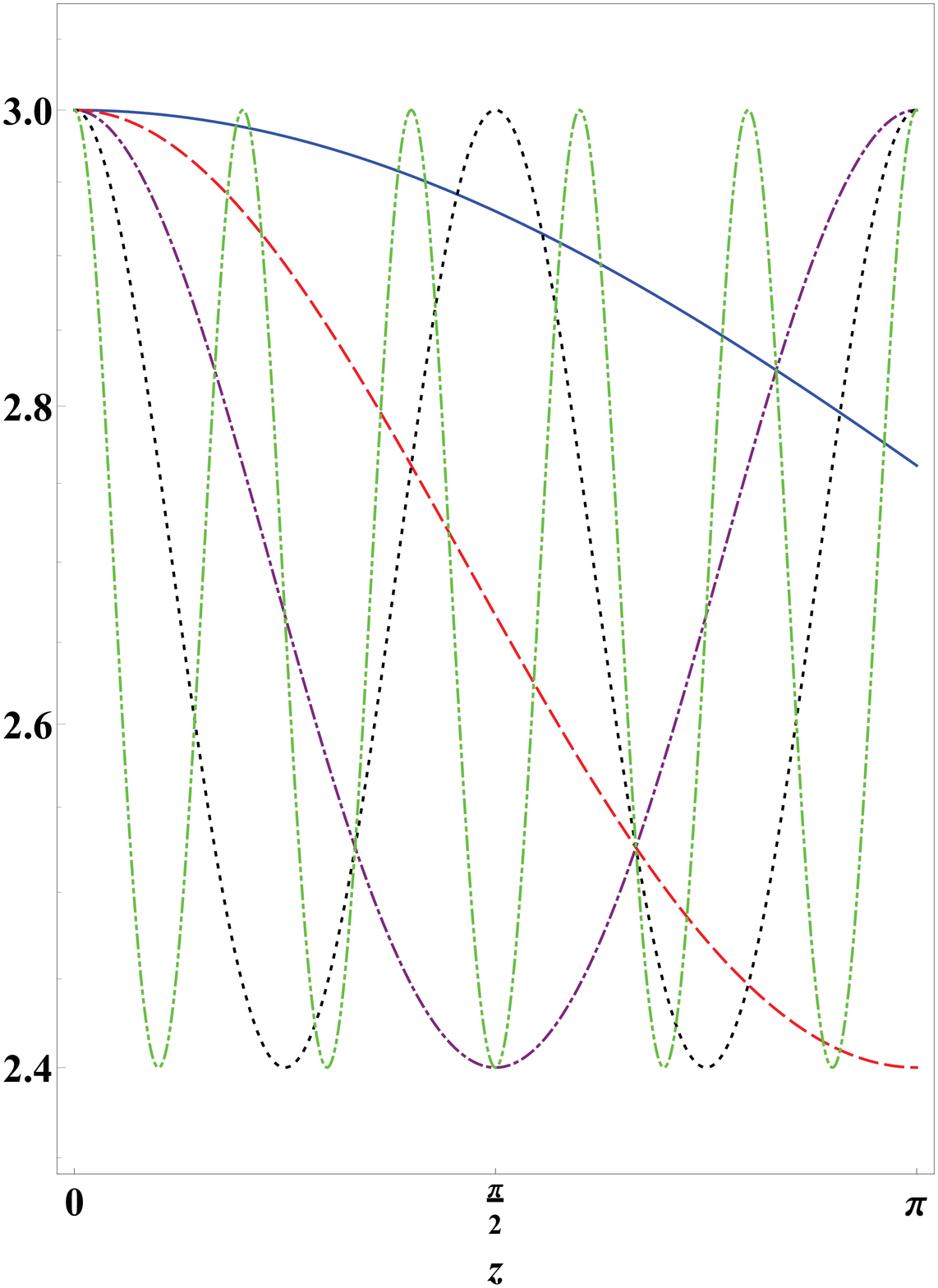}
}
\hspace{0mm} 
\caption{Total energy density (\ref{e:avenertotensre}), divided by $({\cal N}_0/2)^2$, 
versus $z$, at $\varepsilon = \mu + 1 = 2$ and different frequencies:
$\omega/\omega_u = 1/5$ (solid curves), 
$1/2$ (dashed curves),
$1$ (dash-dotted curves),
$2$ (dotted curves)
and $5$ (dash-double-dotted curves),
where $\omega_u$ equals to $\omega_0$ (left panel) or 1 Hz (right panel).
The horizontal axis's units are $h_- \omega_0$ (left panel)
or $h_- \times \text{Hz}$ (right panel).
%Solid curves correspond to $\Lambda_+$, dashed ones -- to $\Lambda_-$.
%The initial state is given by.
}
\label{f:tlsretoten}
\end{figure}

\begin{figure}[t]
\centering
\subfloat[$\omega_0 \not= 0$]{
  \includegraphics[width=0.49\columnwidth]{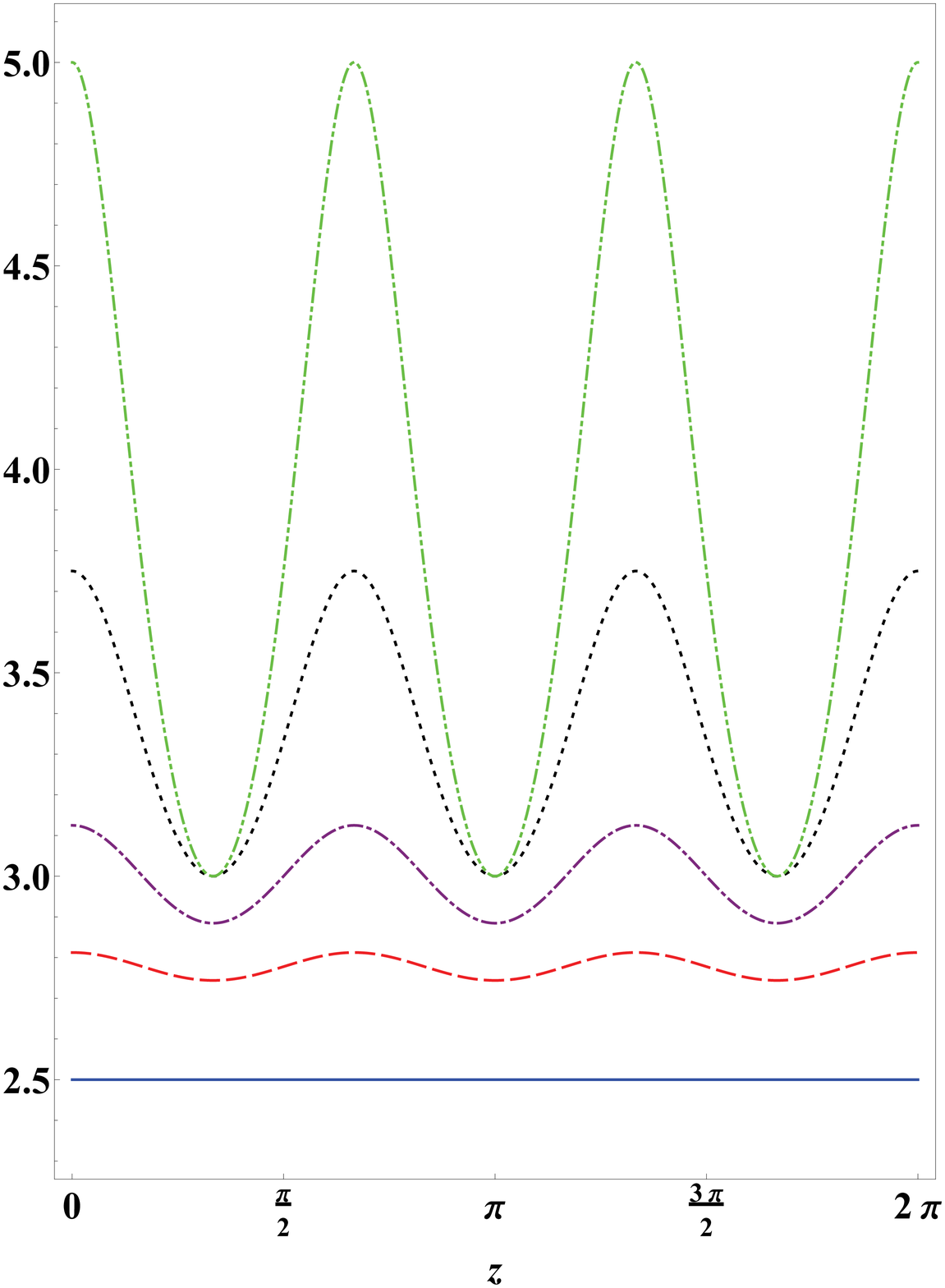}
}
%\hspace{3mm}
\subfloat[$\omega_0 = 0$]{
  \includegraphics[width=0.49\columnwidth]{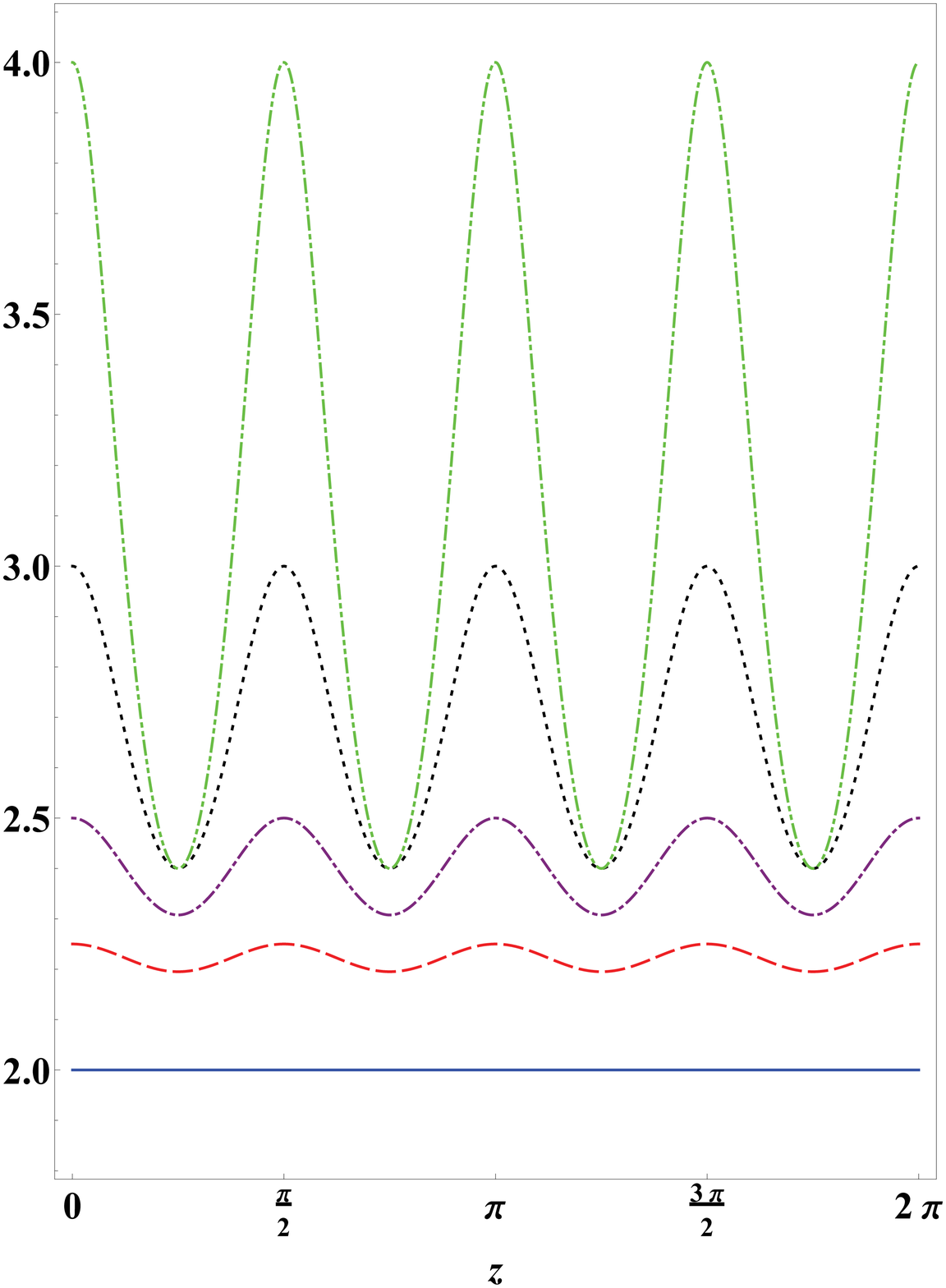}
}
\hspace{0mm} 
\caption{Total energy density (\ref{e:avenertotensre}), divided by $({\cal N}_0/2)^2$, 
versus $z$, at $\omega/\omega_u = 2$, $\mu =1$ and various values of relative permittivity:
$\varepsilon = 1$ (solid curves), 
$3/4$ (dashed curves),
$3/2$ (dash-dotted curves),
$2$ (dotted curves),
and $3$ (dash-double-dotted curves),
where $\omega_u$ equals to $\omega_0$ (left panel) or 1 Hz (right panel).
The horizontal axis's units are $h_- \omega_0$ (left panel)
or $h_- \times \text{Hz}$ (right panel).
%Solid curves correspond to $\Lambda_+$, dashed ones -- to $\Lambda_-$.
%The initial state is given by.
}
\label{f:tlsretoteneps}
\end{figure}

\begin{figure}[t]
\centering
\subfloat[$\omega_0 \not= 0$]{
  \includegraphics[width=0.49\columnwidth]{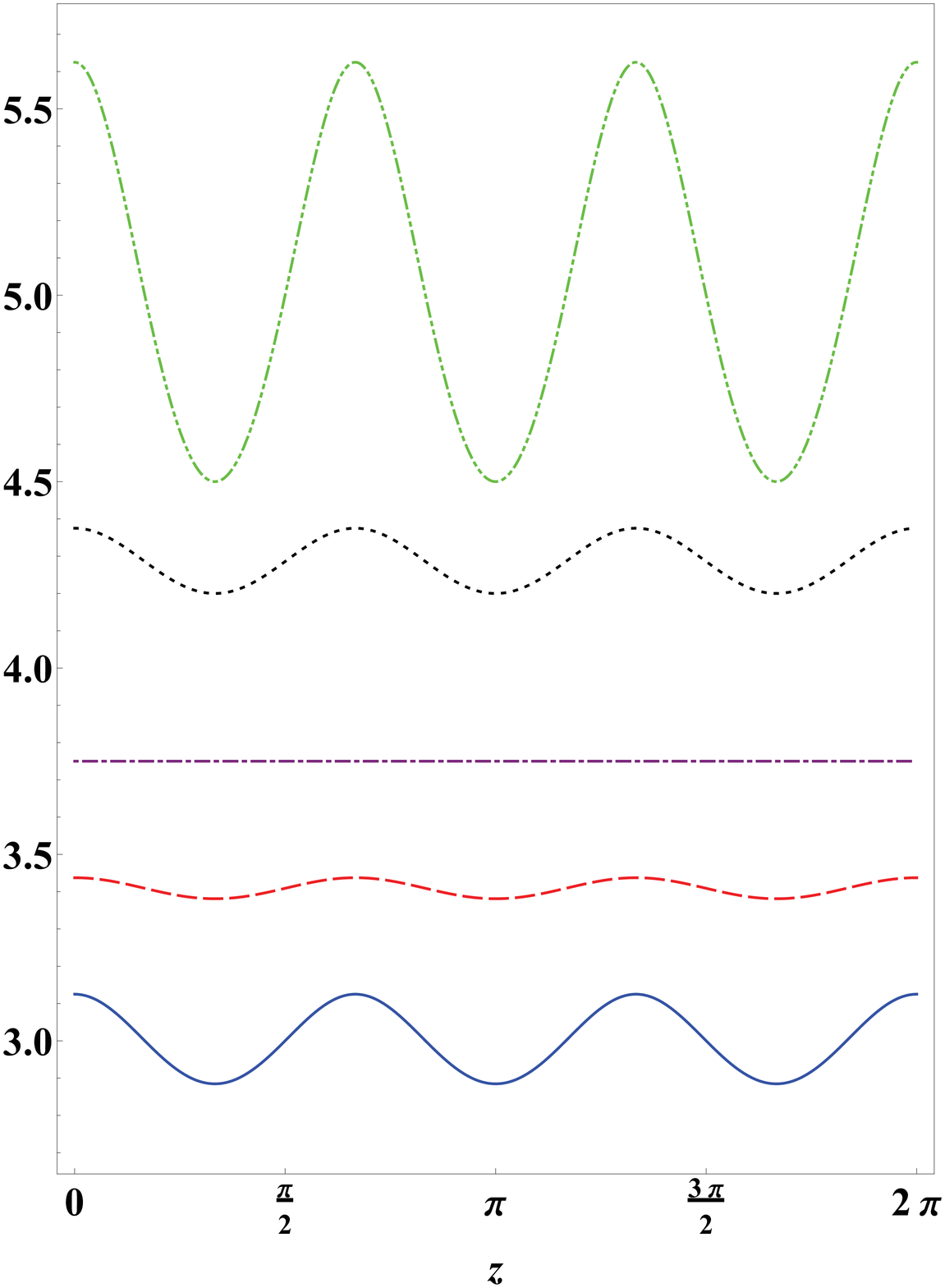}
}
%\hspace{3mm}
\subfloat[$\omega_0 = 0$]{
  \includegraphics[width=0.49\columnwidth]{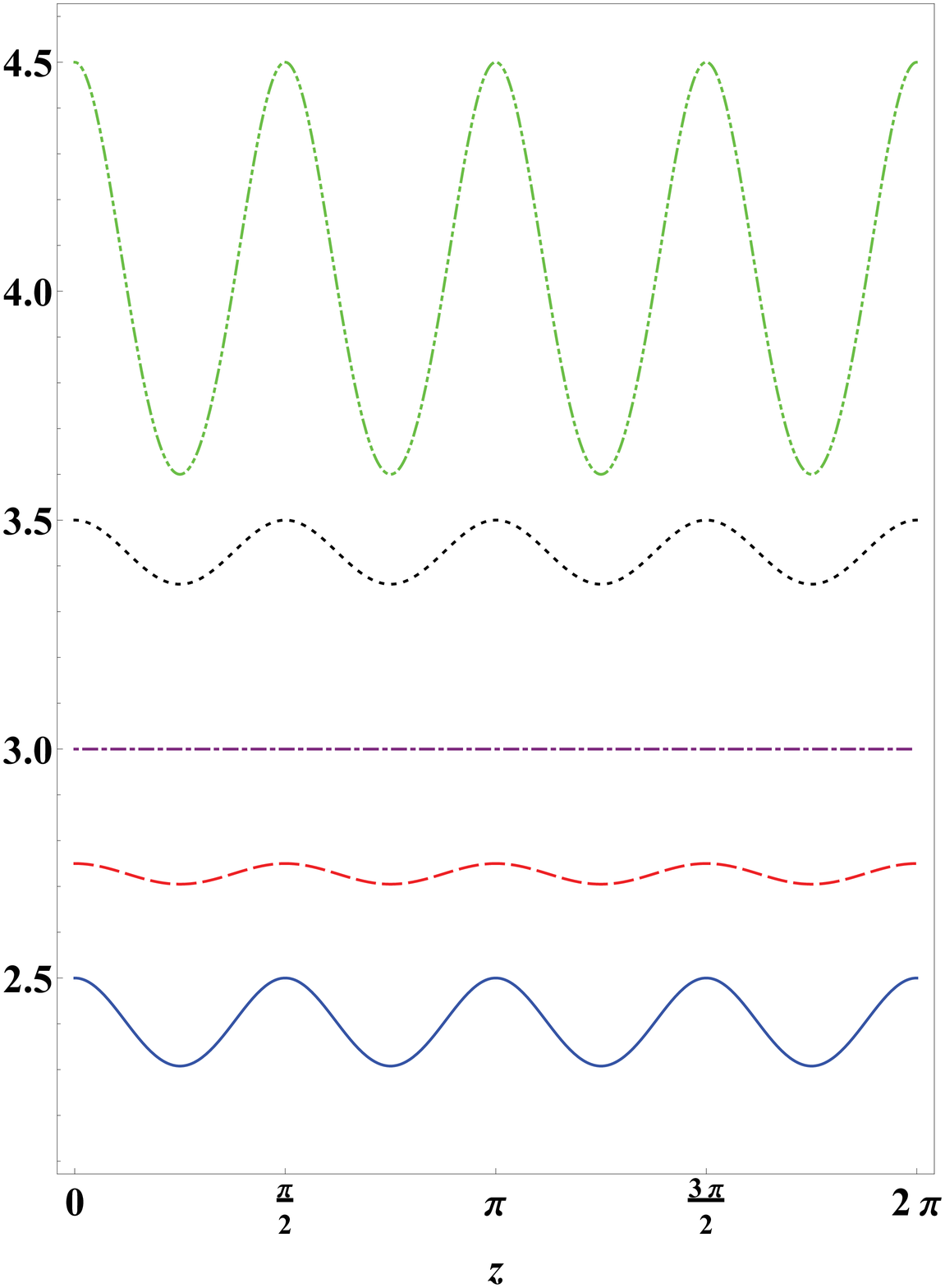}
}
\hspace{0mm} 
\caption{Total energy density (\ref{e:avenertotensre}), divided by $({\cal N}_0/2)^2$, 
versus $z$, at $\omega/\omega_u = 2$, $\varepsilon = 3/2$ and various values of relative permeability:
$\mu = 1$ (solid curves), 
$3/4$ (dashed curves),
$3/2$ (dash-dotted curves),
$2$ (dotted curves),
and $3$ (dash-double-dotted curves),
where $\omega_u$ equals to $\omega_0$ (left panel) or 1 Hz (right panel).
The horizontal axis's units are $h_- \omega_0$ (left panel)
or $h_- \times \text{Hz}$ (right panel).
%Solid curves correspond to $\Lambda_+$, dashed ones -- to $\Lambda_-$.
%The initial state is given by.
}
\label{f:tlsretotenmu}
\end{figure}

%\bc
\textit{(b) Density matrix and averages}.
%\ec
The exact solutions of the master equations for the
density operators of the system
can be found in the form given
by Eqs. (\ref{e:decompom}) and (\ref{e:decomprho}),
where the averages $\avo{\sigma_k}$ and $\av{\sigma_k} $ satisfy Eqs.
(\ref{e:primavoLeqL}) and (\ref{e:primaveqL}).
The latter become, respectively,
\be\lb{e:primavoLeqLRe}
%\begin{matrix}
\frac{1}{2}
\frac{d}{d \t}
\fourcol{
\avo{\sigma_1}}{
\avo{\sigma_2}}{
\avo{\sigma_3}}{
\text{tr} \densnnorm
}
=
\begin{pmatrix}
0 & 0 & \eigv + & \eigv - \\[0.3ex]
0 & 0 & 0 & 0 \\[0.3ex]
- \eigv + &  0 & 0 & 0  \\[0.3ex] 
\eigv - &  0 & 0 & 0
\end{pmatrix}
\!
\fourcol{
\avo{\sigma_1}}{
\avo{\sigma_2}}{
\avo{\sigma_3}}{
\text{tr} \densnnorm
}
,
\ee
and
\be\lb{e:primaveqLRe}
%\begin{matrix}
\frac{1}{2}
\frac{d}{d \t}
\threecol{
\av{\sigma_1}}{
\av{\sigma_2}}{
\av{\sigma_3}}
=
\begin{pmatrix}
\av{\decoO} & 0 & \eigv + \\
0 & \av{\decoO} & 0 \\
- \eigv + &  0 &  \av{\decoO} 
\end{pmatrix}
\!
\threecol{
\av{\sigma_1}}{
\av{\sigma_2}}{
\av{\sigma_3}}
+
\threecol{\eigv -}{0}{0}
,
\ee
where 
\be\lb{e:avGammapLRe}
\av{\decoO}
%\equiv \frac{i}{2} \av{\hamo - \hamo^\dagger}
=
- \eigv - \av{\sigma_1} 
%+ \evim_- \av{\sigma_2}
,
\ee
and
$\eigv{\pm}$ are given in Eq. (\ref{e:lampmhomo}). 

As per usual, 
these equations must be supplemented with the ``initial'' (boundary) conditions
(\ref{e:doini}) where for $\densnnorm_0$ we can choose 
either (\ref{e:rhobloch}) or (\ref{e:rhoident}).
It turns out that the latter is suitable in this case since
the corresponding normalized density matrix (\ref{e:nrhoident})
%has no off-diagonal terms and thus 
describes the classical-type mixture
of two states with equal probabilities to happen.
%Thus, we can assume the initial density matrix given by Eq. (\ref{e:rhoident}).
Using Eqs. (\ref{e:decompom}) and (\ref{e:rhoident}), 
we derive the following conditions for the non-normalized averages:
\be\lb{e:inire}
\avo{\sigma_a}|_{\t=0}
=
0
, \
\text{tr} \densnnorm |_{\t=0} = 2
,
\ee
where $a=1,2,3$.
Correspondingly,
the solution of Eqs. (\ref{e:primavoLeqLRe}) is
\bse
\ba
\avo{\sigma_1} 
&=&
\frac{h_-}{n} \Sin{}{\pcz \t}
,\\
\avo{\sigma_2} 
&=&
0
,\\
\avo{\sigma_3} 
&=&
-
\frac{1}{2 n^2}
F_+ (\t)
,\\
\text{tr} \densnnorm
&=&
\frac{1}{2 n^2}
F_- (\t)
,
\ea
\ese
where 
\be
\pcz
=
2 n  \om
,
\ee
and
we have denoted the functions
\ba
F_+ (\t) 
&=&
h_+ h_- (1 - \Cos{}{\pcz \t})
,\nn\\
F_- (\t) 
&=&
h_+^2 - h_-^2
\Cos{}{\pcz \t}
,\nn
\ea
where the former 
is non-negative at $\varepsilon \geqslant \mu \geqslant 1$,
and the latter is always positive for the materials with positive permittivity and permeability.
This solution indicates a presence of a stationary wave, therefore it contains
interference terms described by the linear combinations of sine or cosine functions.
%A single mode exists because the cross section is uniform.
% and there is no dependence on transverse coordinates of all of the fields. 

Consequently, the normalized expectation values are
\bse\lb{e:avtlsre}
\ba
\av{\sigma_1} 
&=&
2 n h_-
\Sin{}{\pcz \t}/
F_- (z)
,\\
\av{\sigma_2} 
&=&
0
,\\
\av{\sigma_3} 
&=&
-
F_+ (\t) / F_- (\t)
.
\ea
\ese

It is clear from these expressions that our statistical ensemble of wave modes
has a periodic structure along $\t$-direction, with the period equal to 
\be\lb{e:tlsreperiod}
T_z
=
2\pi / |\pcz|
=
\pi / (n |\om|)
,
\ee
so its behavior strongly depends on the wave frequency $\omega$,
see Fig. \ref{f:tlsreperiod}.
%There exists one exceptional case, though: 
When $\omega \not= 0$ the dependence on $\t$ can still disappear -- when $\pcz \to 0$,
which is equivalent to $\omega \to \omega_0$.
Thus, $\omega_0$ is a critical frequency at which the oscillations of the statistical
averages
become suppressed.

%\bc
\textit{(c) Observables: wave energy}.
%\ec
Once we know the exact solution for density matrix, we have all the
information about the probability weights of every mode that forms the beam,
therefore,
a number of energy-related averages, which have been defined in Sec. \ref{s-nhgen-ave},
can be easily computed.
Due to real-valued permittivity and permeability, some formulas of Sec. \ref{s-nhgen-ave}
get simplified:
\ba
&&
\wenbar_\bot
=
\frac{{\cal N}^2_0}{
%32 \pi 
8\omega}
\av{
{\cal L}_+
\hamod
}
%\nn\\&&\quad \quad
=
\frac{{\cal N}^2_0}{2}
G (z)
,\\&&
\wenbar_z
=
\frac{{\cal N}^2_0}{
%32 \pi 
8
\omega}
\av{
{\cal L}_-
\hamod
}
=
\frac{{\cal N}^2_0 \omega^2_0}{2 \omega^2}
%\frac{\omega^2_0}{\omega^2}
G(z)
,\\&&
\wenbar_\text{tot} = \wenbar_\bot + \wenbar_z
=
\frac{{\cal N}^2_0}{
%32 \pi 
4}
\left\langle 
\pDer \omega
\hamod
\right\rangle
\nn\\&&\qquad
=
\frac{{\cal N}^2_0}{2}
\left(
1 +
%\left(
\frac{\omega_0^2}{\omega^2}
%\right)^2
\right)
G (z)
,
\lb{e:avenertotensre}
\\&&
\Xi_\wen
\equiv
\frac{
\wenbar_z}{\wenbar_\bot}
=
\frac{
\av{
{\cal L}_-
\hamod
}
}{
\av{
{\cal L}_+
\hamod
}
}
=
\frac{\omega^2_0}{\omega^2}
,\\&&
\Xi_\bot
\equiv
\frac{\wenbar_\bot}{ \wenbar_\text{tot} }
= 
1 - \frac{\wenbar_z}{\wenbar_\text{tot} }
=
\left(
1 +
%\left(
\frac{\omega_0^2}{\omega^2}
%\right)^2
\right)^{-1}
,
\ea 
where
\be
G (z) =
\frac{h_+}{4}
\left(
1+
\frac{h_-}{h_+} \av{\sigma_3}
\right)
=
\frac{ n^2 h_+
}{
F_- (z)}
,
\ee
and
%\bse\lb{e:vwenbaremre}
\ba
&&
\vwenbar_\bot^{(e)}
=
{\cal N}^2_0
\av{\hat e}
=
\frac{{\cal N}^2_0}{2} 
G_+ (z) 
,\lb{e:vwenbarere}\\&&
\vwenbar_\bot^{(m)}
=
{\cal N}^2_0
\av{\hat g}
=
\frac{{\cal N}^2_0}{2} 
G_- (z) 
,
\lb{e:vwenbarmre}
\\&&
\Xi_{\vwen}
=
\frac{\vwenbar_\bot^{(e)}}{\vwenbar_\bot^{(m)}}
%= \frac{\av{\hat e}}{\av{\hat g}}
%= \frac{G_+ (z) }{G_- (z) }
=
\frac{
\mu h_+ + \varepsilon h_-
\Cos{}{\pcz \t}
}{
\varepsilon h_+ - \mu h_-
\Cos{}{\pcz \t}
}
,
\ea
where
\be
G_\pm (z) =
1 \pm
\av{\sigma_3}
=
1 \mp
\frac{F_+ (\t)}{F_- (\t)}
,
\ee
and
\be\lb{e:tpowerre}
\tpowerbar
=
\frac{{\cal N}^2_0}{
%16 \pi
4}
\av{\sigma_2}
=
0
,
\ee
where the averages are given by Eqs. (\ref{e:avtlsre}),
and we have used the identity
\[
\pDer \omega
\hamod
=
\frac{1}{\om}
\left(
1 + 
%\left(
\frac{\omega_0^2}{\omega^2}
%\right)^2
\right)
\hamod
,
\]
which can be easily derived from Eqs. (\ref{e:hamtodIItls}) and (\ref{e:lamtlsRe2}).

From these expressions,
one can immediately spot a few universal features of the system.
For example, energy density (\ref{e:avenertotensre}), as well as its parts,
are positive functions if medium's permittivity and permeability are positive values. 
These functions are oscillatory but become constant when $\varepsilon = \mu$ or $\pcz = 0$.
The former condition defines media in which the oscillations are suppressed, therefore,
the wave propagation is similar to the one in the vacuum.
The latter condition is discussed above, after Eq. (\ref{e:tlsreperiod}).
Further,
if $\omega_0 \to 0$ then all energy gets concentrated in the transverse
part $\wenbar_\bot$.
When $\omega_0 \not=0$, the energy tends to move 
to the transverse part at large frequencies $\omega$ 
and to the longitudinal part $\wenbar_z$ at small frequencies.

The example profiles of the energy-related functions of $\t$, at different values
of $\varepsilon$, $\mu$ and $\omega$, are given
in Figs. \ref{f:tlsretoten}-\ref{f:tlsretotenmu}.\\

%\bc
\textit{(d) Entropy}.
%\ec
In this case, the ``initial'' (boundary) values of the notions of entropy,
which were
introduced in Sec. \ref{s-nhgen-entr},
are:
\be
\left.
S_\text{vN}
\right|_{z=0} = \ln 2,
\
\left.
S_\text{NH}
\right|_{z=0}
= 0,
\ee
according to Eqs. (\ref{e:SVNnH}), (\ref{e:SVNnH2}), (\ref{e:decompom}), (\ref{e:decomprho})
and (\ref{e:inire}).
The Gibbs-von-Neumann entropy (\ref{e:SVNnH})
for our density matrix can be computed as (in units $k_B =1$):
\bw
\be\lb{e:SVNnHre}
%k_B^{-1}
S_\text{vN}
=
\ln{\!(4 n^2)}
-\frac{1}{2}
\text{tr}
\left\{
\left(
\hat I
+
%\av{\sigma_1}\,
2 n h_-
\frac{\Sin{}{\pcz \t}}
{F_- (z)} 
\hat\sigma_1
%+ \av{\sigma_3}\, 
-
\frac{F_+ (\t)}{F_- (\t)}
\hat\sigma_3
\right)
\ln{\!
\left[
%\text{tr} \densnnorm\, 
%\frac{1}{2 n^2}
F_- (\t)
\hat I
+
%\avo{\sigma_1}\,
2 n h_- \Sin{}{\pcz \t} 
\hat\sigma_1
%+ \avo{\sigma_3}\, 
-
F_+ (\t)
\hat\sigma_3
\right]
}
\right\}
,
\ee
\ew
whereas the NH entropy (\ref{e:SVNnH2})
can be derived from the relation
\be\lb{e:diffentre}
S_\text{NH}
=
S_{\rm vN}
-
%k_{\rm B}
\ln{[
F_- (\t)]}
+
%k_{\rm B}
\ln{\!(2 n^2)}
,
\ee
which follows from Eq. (\ref{e:diffentr})

\begin{figure}[t]
\centering
\subfloat[$\omega_0 \not= 0$]{
  \includegraphics[width=0.49\columnwidth]{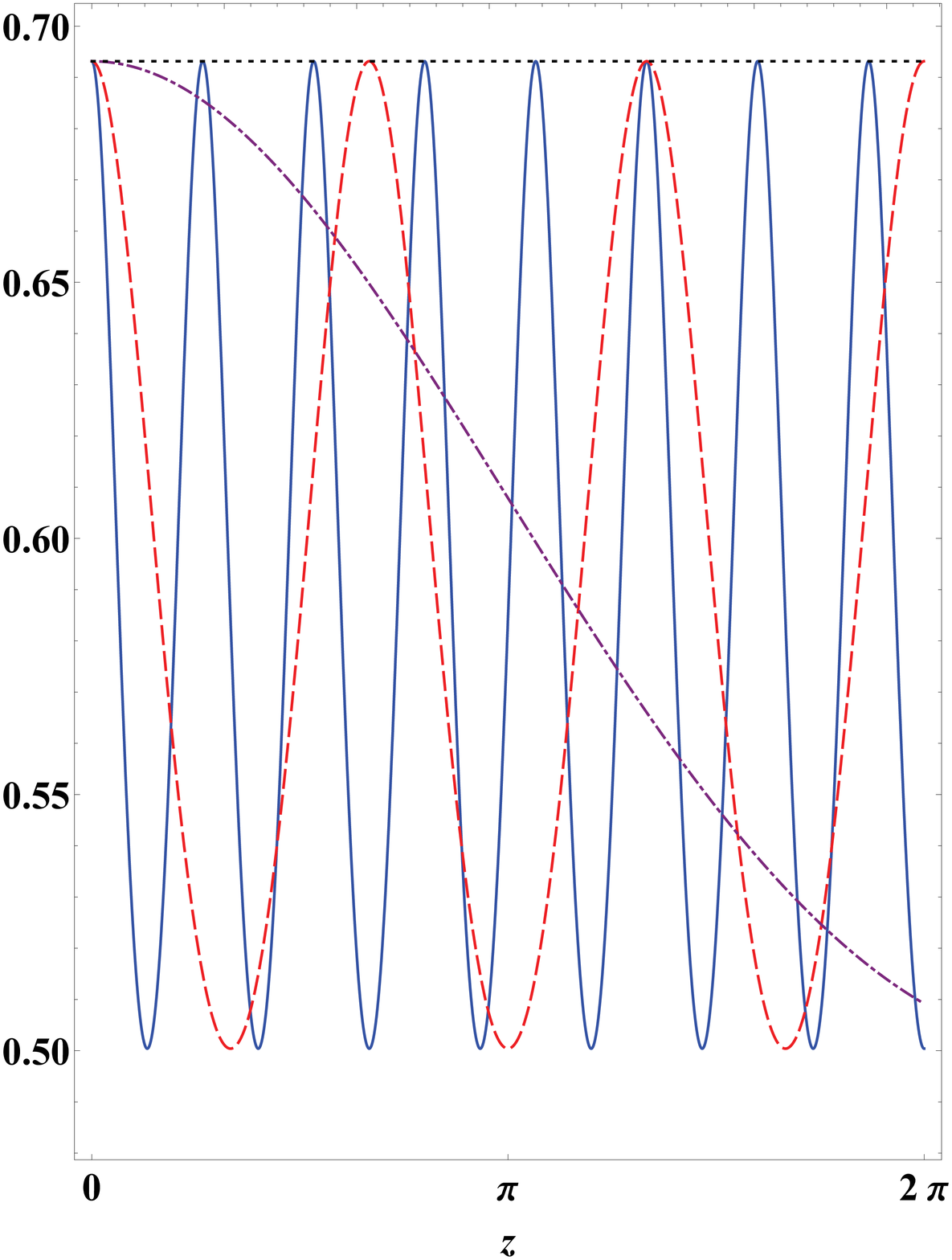}
}
%\hspace{3mm}
\subfloat[$\omega_0 = 0$]{
  \includegraphics[width=0.49\columnwidth]{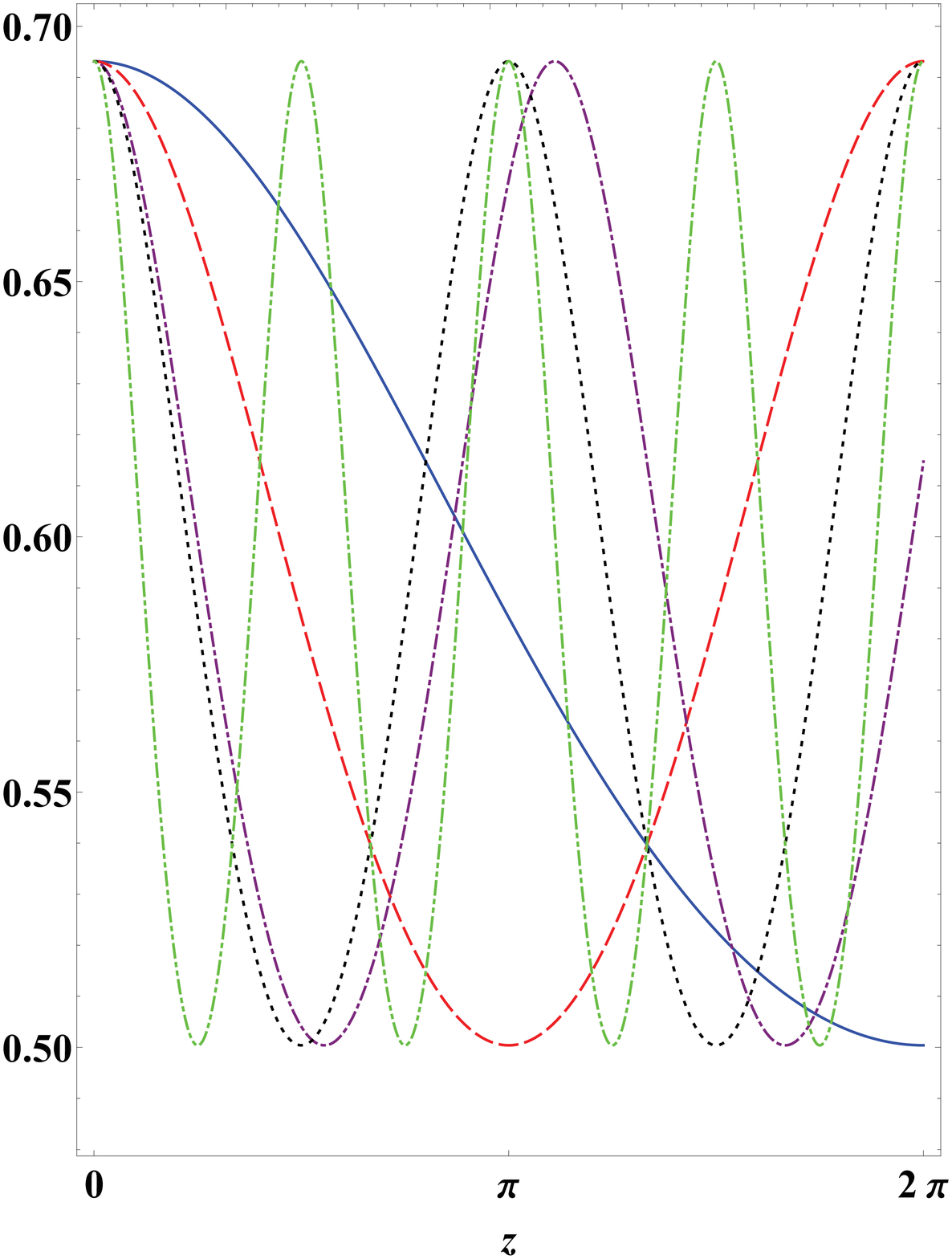}
}
\hspace{0mm} 
\caption{Gibbs-von-Neumann entropy (\ref{e:SVNnHre})
%, in units of $k_B$,
%(\ref{e:diffentre})
versus $z$, at $\varepsilon = \mu + 1 = 2$ and different frequencies:
$\omega/\omega_u = 1/4$ (solid curves), 
$1/2$ (dashed curves),
$0.9$ (dash-dotted curves),
$1$ (dotted curves)
and $2$ (dash-double-dotted curve, right panel only),
where $\omega_u$ equals to $\omega_0$ (left panel) or 1 Hz (right panel).
The horizontal axis's units are $h_- \omega_0$ (left panel)
or $h_- \times \text{Hz}$ (right panel).
%Solid curves correspond to $\Lambda_+$, dashed ones -- to $\Lambda_-$.
%The initial state is given by.
}
\label{f:tlsresvn}
\end{figure}

The behavior of the Gibbs-von-Neumann entropy at different values
of frequency $\omega$ is
illustrated in Fig. \ref{f:tlsresvn}.
One can see that this entropy is oscillating between its initial value, $\ln 2$,
and value $1/2$, and the period of oscillations depends on the frequency,
as expected from Eq. (\ref{e:tlsreperiod}).
Therefore, for the case $\omega_0 \not=0$ it suffices to consider the plots
with $\omega \leqslant \omega_0 $ since this entropy is invariant
under the transformation $\tilde\omega \to 1/\tilde\omega$.

\scn{Conclusion}{s-con}

In this paper, using the formal analogy between the
\schrod~equation and a certain class of Maxwell equations, 
we have generalized the theory of EM wave's propagation in dielectric linear media
-- in order to 
be able to describe not only separate wave modes (or their linear superpositions) 
but also the statistical ensembles of modes, referred as mixed states in quantum mechanics.

It turns out that the Hamiltonians, which govern the dynamics of
such ensembles, are in general not just pseudo-Hermitian 
%or parity-time-symmetric
but essentially non-Hermitian and thus require a
special systematic treatment.
Using the density operator approach for general
non-Hermitian Hamiltonians developed in our earlier works,
we have demonstrated that the non-Hermitian terms play an
important role in the physics of wave propagation.

The proposed approach applies to a large class of dielectric media and
nanoscale photonic and plasmonic materials and wave-guiding devices,
where it provides a tool to construct and study different models,
as well as to derive the observables of different kinds:
correlation functions (Sec. \ref{s-nhgen-corr}), entropy (Sec. \ref{s-nhgen-entr}), energy density and transmitted power (Sec. \ref{s-nhgen-ave}), \textit{etc}.  
For instance, the introduced notions of entropy are important for estimating
the degree of statistical uncertainty and chaos in a given system,
whereas the statistically averaged values of energy density and transmitted power
are helpful for describing the dissipative effects in the system
due to interaction of different modes, which lead to energy and information loss.
The method sheds light upon various quantum-statistical effects that can occur,
such as the additional corrections to the conservation equation
for the transmitted power, which arise due to the quantum exchange of energy
between the medium and electric and magnetic components of wave modes, see Sec. \ref{s-nhgen-ave}.
Another effect, demonstrated by one of our examples, in Sec. \ref{s-tls-rea},
reveals that quantum-statistical corrections can make the wave's propagation essentially
dispersive, even if the media itself has frequency-independent 
permittivity and permeability. 
That example also
demonstrates that non-Hermitian terms in Hamiltonians do not always lead to energy loss
but can describe, under certain conditions,
the oscillating behavior of statistical uncertainty
in the system which can be related to certain kinds of noise.

This results in a consistent and thorough understanding of whether and how one can control the 
dissipative effects in different dielectric media, which lead to decoherence and energy and information loss
during propagation of EM waves.
The control over these effects is 
especially important for the development of 
the next generation of quantum electromagnetic devices, including those
which use the quantum interference of multimode EM beams
in order to improve
the sensitivity and non-invasivity of measurements,  
quantum amplifiers and radars being just some examples \cite{cvk99,gat00,llo08,tan08,lop13,bar15}.
For instance, the uncontrolled spontaneous transition of pure modes into 
statistical ensembles during beam's propagation would inevitably result in an increase of statistical uncertainty
and hence lead to higher degrees of dissipation and noise.
Further studies of such quantum-statistical effects is a fruitful direction of future research.

Last but not least, one can use this approach both ways:
it also provides a methodology of how one can use electromagnetic phenomena
%in dielectric media 
for experimental testing of the heuristic 
concepts and ideas of the non-Hermitian formalism \textit{per se}, such as non-normalized and normalized density
operators, master equations with anti-commutators, nonlinear and nonlocal terms, different notions
of entropy, to mention just a few examples.

\begin{acknowledgments}
This paper is based on seminars given during my visit to Aston University and
Aston Institute of Photonic Technologies (AIPT), Birmingham, UK, in November 2015.
I would like to thank Michael ``Misha'' Sumetsky for the hospitality during my visit,
as well as I acknowledge thought-provoking discussions with him, Mykhaylo Dubov, Igor Yurkevich,
and other members of AIPT and participants of the seminars.
%Fruitful discussions with participants of the 9th International Kharkiv Symposium on Physics and Engineering of Microwaves, Millimeter and Submillimeter Waves MSMW (20-24 June, 2016, Kharkiv, Ukraine), where parts of this work have been presented, 
%are acknowledged.
Proofreading of the
manuscript by P. Stannard is greatly acknowledged.
% as well.
Both this work and my visit to AIPT are supported by the National Research Foundation of South 
Africa 
under Grants 
%95965, 
98083 and 98892.
%This work is based on the research supported wholly / in part by the National Research Foundation of South Africa (Grant Numbers xxx, yyy, and zzz)
\end{acknowledgments}

\appendix

\scn{Operator algebra and observables for EM wave modes}{a:op} 

In case of a two-dimensional transverse space,
one can define the SU(2) algebra
by using the vector product with the basis vector along the
longitudinal direction, $\vee z$.
Indeed, when applied to a 2D vector, the operator $\vee z \times$
acts as 
% a product with 
the imaginary unit,
\be
\left(\vee z \times \right)^\dagger = 
- \vee z \times 
, \ \
\vee z \times \vee z \times = - \hat I
,
\ee
which means that 
%the operator  $\hat Z \equiv - i \vee z \times$
it 
is anti-Hermitian, anti-involutary and anti-unitary operator in 
%two-dimensional Hilbert 
the space
of two-dimensional vectors. 
Therefore, using this operator one can define the following
set of
Pauli matrices
\bse
\ba
&&
\hat\sigma_1
=
-i \twomat{0}{\vee z \times}{\vee z \times}{0}
, 
\\&&
\hat\sigma_2
=
\twomat{0}{-\vee z \times}{\vee z \times}{0}
, 
\\&&
\hat\sigma_3
=
\twomat{\hat I}{0}{0}{- \hat I}
%\twomat{- i \vee z \times}{0}{0}{i \vee z \times}
,
\ea
\ese
which are Hermitian, involutary, unitary, traceless,
of a unit determinant, and
satisfy the commutation relations
\ba
&&
\left\{\hat\sigma_a, \hat\sigma_b\right\} = 2 \delta_{a b} \hat I
,
\\&&
\left[\hat\sigma_a, \hat\sigma_b\right] = 2 i \sum\limits^3_{c=1}\epsilon_{a b c} \hat\sigma_c
,
\ea
where $\delta_{a b}$ and $\epsilon_{a b c}$ are the Kronecker and Levi-Civita symbols, respectively.
Besides, 
\ba
&&
\hat\sigma_a \hat\sigma_b
=
i \sum\limits^3_{c=1}\epsilon_{a b c} \hat\sigma_c
+
\delta_{a b} \hat{\cal I}
,\\
&&
\hat\sigma_1^2 = \hat\sigma_2^2 = \hat\sigma_3^2 = -i \hat\sigma_1 \hat\sigma_2 \hat\sigma_3 
= \hat{\cal{I}}
= \twomat{\hat I}{~0}{0}{~\hat I}
.
~~~
\ea
The expectation values of these Pauli matrices,
\be
\left\langle \sigma_a \right\rangle_\Psi \equiv 
\left\langle \Psi | \hat\sigma_a | \Psi \right\rangle
,
\ee
have a physical interpretation in terms 
of energies associated with the EM wave:
using (\ref{e:wfnorm}) and (\ref{e:ketnorm}) one obtains
%\bw
\bse
\ba
\left\langle \sigma_1 \right\rangle_\Psi
&=&
\frac{i}{{\cal N}^2} 
%{\rm Im}
\int d x d y 
\left[
(\vee z\times\veeft^*)\cdot\vemft
-
\text{c.c.}
\right]
\nn\\&=&
%=
\frac{i}{{\cal N}^2}
%{\rm Im}
\int d x d y 
\left(
E^*_{[x} H_{y]}
-
\text{c.c.}
\right)
,
\\
\left\langle \sigma_2 \right\rangle_\Psi
&=&
\frac{1}{{\cal N}^2} 
%{\rm Im}
\int d x d y 
\left[
(\vee z\times\veeft^*)\cdot\vemft
+
\text{c.c.}
\right]
\nn\\&=&
\frac{1}{{\cal N}^2}
%{\rm Im}
\int d x d y 
\left(
E^*_{[x} H_{y]}
+
\text{c.c.}
\right)
=
%\frac{16 \pi
%}{c}
\frac{
%16 \pi 
4
\tpower}{{\cal N}^2}
,~~~
\lb{e:onemodetpower}
\\
\left\langle \sigma_3 \right\rangle_\Psi
&=&
\frac{1}{{\cal N}^2}
\int d x d y 
\left(
%\veeft\cdot\veeft^*  - \vemft\cdot\vemft^*
|\veeft|^2  - |\vemft|^2
\right)
%\nn\\&=& \frac{1}{{\cal N}^2} \int d x d y \left( |E_x|^2 + |E_y|^2 - |H_x|^2 - |H_y|^2 \right)
,
\ea
\ese
%\ew
where the integrals are taken over waveguide's effective cross-section,
and we use the notation 
$A_{[j} B_{m]} = A_j B_m - A_m B_j$.
Here  $\tpower$ is 
the transmitted power carried by an EM wave mode.
One can  also borrow the notations from theory of two-level systems
and introduce the operators
\bse
\ba
&&
\hat g
\equiv
\left|g \right\rangle\!\left\langle g \right|
\equiv
\frac{1}{2} 
\left(
\hat{\cal{I}}
-
\hat\sigma_3
\right)
=
\twomat{0}{~0}{0}{~\hat I}
,
\\&&
\hat e
\equiv
\left|e \right\rangle\!\left\langle e \right|
\equiv
\frac{1}{2} 
\left(
\hat{\cal{I}}
+
\hat\sigma_3
\right)
=
\twomat{\hat I}{~0}{0}{~0}
,
\ea
\ese
where
\be
\left|g \right\rangle 
\equiv
\twocol{0}{\hat I}
, \
\left|e \right\rangle 
\equiv
\twocol{\hat I}{0}
,
\ee
such that
\bse
\ba
&&
\left\langle \hat g \right\rangle_\Psi
=
\frac{1}{{\cal N}^2}
\int 
d x d y 
|\vemft|^2
,
\\&&
\left\langle \hat e \right\rangle_\Psi
=
1 - 
\left\langle \hat g \right\rangle_\Psi
=
\frac{1}{{\cal N}^2}
\int
d x d y  
|\veeft|^2
%\\&& \left\langle \hat g \right\rangle_\Psi + \left\langle \hat e \right\rangle_\Psi = 1
,
\ea
\ese
where letters $g$ and $e$ indicate the ``ground'' and ``excited'' states, respectively.
From the viewpoint of electrodynamics, the ratio
\be
%\Xi \equiv
\left\langle \hat e \right\rangle_\Psi / 
\left\langle \hat g \right\rangle_\Psi
=
\left[
\frac{1}{
\left\langle \hat e \right\rangle_\Psi
}
- 1
\right]^{-1}
\ee
describes how much of wave's ``pure'' energy (in absence of a medium) would be stored in the electric component
as compared to the magnetic one.

\scn{Bloch-sphere parametrization for EM wave modes}{a:bs} 

Any Hermitian operator $\hat\varrho$, which
has trace one and idempotency property $\hat\varrho^2 = \hat\varrho$,
can
be parametrized using the Bloch sphere:
\ba
\hat\varrho
&=&
\left|\Psi_0 \right\rangle\!\left\langle \Psi_0 \right|
=
%\bcs
%\frac{1}{2} \twomat{1+\tilde\varrho}{\varrho_1 + i \varrho_2}{\varrho_1 - i \varrho_2}{1-\tilde\varrho} 
%\twomat{\frac{1}{2}(1-\bar\varrho)}{\varrho_1 + i \varrho_2}{\varrho_1 - i \varrho_2}{\frac{1}{2}(1+\bar\varrho)} 
%{0<\varrho_1<1 \choose{\varrho_2^2 < \varrho_1 - \varrho_1^2}} 0<\varrho_1<1
\twomat
{\Sin{2}{\theta_0/2}}
{\frac{1}{2}\text{e}^{i \phi_0}\sin\theta_0}
{\frac{1}{2}\text{e}^{-i \phi_0}\sin\theta_0}
{\Cos{2}{\theta_0/2}}
\nn\\&
=&
\frac{1}{2}
\left[
\hat I
+
\sin{\theta_0}
(
\cos{\phi_0} \,
\hat\sigma_1
-
%\sin{\theta_0}
\sin{\phi_0} \,
\hat\sigma_2
)
%\nn\\&&
-
%\frac{1}{2}
\cos{\theta_0} 
\hat\sigma_3
%\frac{1}{2}
\right]
,
\nn\\
\lb{e:a-rhobloch}
\ea
where
\ba
&&
\left|\Psi_0 \right\rangle
=
\cos{(\theta_0/2)} 
\left|g \right\rangle
+
\text{e}^{i \phi_0}
\sin{(\theta_0/2)} 
\left|e \right\rangle
,
\\&&
\left|g \right\rangle 
=
\twocol 0 1
, \ \
\left|e \right\rangle 
=
\twocol 1 0
,
\ea
and
$0 \leqslant \theta_0 \leqslant \pi$
and 
$0 \leqslant \phi_0 < 2 \pi$.
Matrix (\ref{e:a-rhobloch}) has the eigenvalues $0$ and $1$,
%we have denoted 
%$\tilde\varrho = \pm\sqrt{1 - \varrho_1^2- \varrho_2^2}$ and assumed that all $\varrho$'s are real-valued.
%Here the cases when $\varrho_1 = 0, 1,
therefore, one of its special cases 
%of a pure state's density matrix
would be the basis states
\be\lb{e:psbas}
\baa{ll}
\left\{\theta_0 = \pi, \ \phi_0 = 0\right\}: &
\hat\varrho^{(e)} = \twomat 1 0 0 0 
=
\hat e
,\\
\left\{\theta_0 = 0, \ \phi_0 = 0\right\}: &
\hat\varrho^{(m)} = \twomat 0 0 0 1 
=
\hat g
,
\eaa
\ee
where 
%we have used TLS notations
\bse
\ba
&&
\hat g
\equiv
\left|g \right\rangle\!\left\langle g \right|
\equiv
\frac{1}{2} 
\left(
\hat{I}
-
\hat\sigma_3
\right)
,\\&&
\hat e
\equiv
\left|e \right\rangle\!\left\langle e \right|
\equiv
\frac{1}{2} 
\left(
\hat{I}
+
\hat\sigma_3
\right)
,
\ea
\ese
which physical meaning 
%of basis states
is clear from Appendix \ref{a:op} and Eqs. (\ref{e:vwenbare}) and (\ref{e:vwenbarm}):
% and (\ref{e:vwenbarm}):
matrices $\hat\varrho^{(e)}$ and $\hat\varrho^{(m)} $ describe the states during wave's
propagation when the 
wave's ``medium-independent'' energy (as if the medium were absent) is stored mostly inside the electrical and magnetic field
components, respectively.
Consequently, we have
\bse
\ba
&&
\left\langle \hat\varrho^{(e)} \right\rangle_\Psi
=
1 - 
\left\langle \hat\varrho^{(m)} \right\rangle_\Psi
=
\frac{1}{{\cal N}^2}
\int
d x d y  
|\veeft|^2
%\\&& \left\langle \hat g \right\rangle_\Psi + \left\langle \hat e \right\rangle_\Psi = 1
,~~~\\&&
\left\langle \hat\varrho^{(m)} \right\rangle_\Psi
=
\frac{1}{{\cal N}^2}
\int 
d x d y 
|\vemft|^2
,
\ea
\ese
where the Appendix \ref{a:op}'s notations are used.
Another example of a pure-state matrix are the following superpositions of basis states:
\ba
\baa{ll}
\left\{\theta_0 = \frac{\pi}{2}, \ \phi_0 = 0\right\}: &
\hat\varrho^{(+)} = \frac{1}{2}\twomat 1 1 1 1 
=
|+ \rangle \langle + |
,\\
\left\{\theta_0 = -\frac{\pi}{2}, \ \phi_0 = 0\right\}: &
\hat\varrho^{(-)} = \frac{1}{2}\twomat{~1}{-1}{-1}{~1}
=
|- \rangle \langle - |
,
\eaa
\ea
where
\ba
|\pm \rangle \langle \pm |
&=&
\frac{1}{2}
\left[
|g\rangle\langle g|
+
|e\rangle\langle e|
\pm
\left(
|g\rangle\langle e|
+
|e\rangle\langle g|
\right)
\right]
\nn\\&
=&
%\frac{1}{2}\sum\limits_{\{j,k\}=\{g,e\}} |j\rangle\langle k|
%\nn\\&=&
\frac{1}{2}
\left(
\hat I
\pm
\hat\sigma_1
\right)
,
\\
|\pm \rangle
&=&
\frac{1}{\sqrt{2}}
\twocol{1}{\pm 1}
=
\frac{1}{\sqrt{2}}
\left(
|e\rangle \pm |g\rangle
\right)
,
\ea
which represent the states when the 
wave's medium-independent energy is distributed between the electrical and magnetic field
components, as one can see by computing the corresponding averages with respect
to a state vector $| \Psi \rangle$:
\be
\left\langle 
\hat\varrho^{(\pm)}
\right\rangle_\Psi
=
1
\pm
\frac{1}{{\cal N}^2}
\int 
d x d y 
\left(
\veeft^*
\cdot
\vemft
+
\text{c.c.}
\right)
.
\ee

%\newpage 
%\bw

\end{document}